\documentclass[12pt,preprint]{aastex}

\def\ne{n$_{\rm e}$\/}

\def\rfe{R$_{\rm FeII}$}
\def\feiiq{\rm Fe{\sc ii }$\lambda$4570\/}

\def\msol{M$_\odot$\/}

\def\rg{$R_{\rm G}$}
\def\rsg{$R_{\rm SG}$}
\def\ltsima{$\; \buildrel < \over \sim \;$}
\def\simlt{\lower.5ex\hbox{\ltsima}}            
\def\gtsima{$\; \buildrel > \over \sim \;$}

\def\simgt{\lower.5ex\hbox{\gtsima}}            
 
\def\a{$\alpha$}

\def\lya{{ Ly}$\alpha$}
\def\civ{{\sc{Civ}}$\lambda$1549\/}
\def\civnc{{\sc{Civ}}$\lambda$1549$_{\rm NC}$\/}
\def\civbc{{\sc{Civ}}$\lambda$1549$_{\rm BC}$\/}

\def\cm3{cm$^{-3}$\/}
\def\hb{{\sc{H}}$\beta$\/}

\def\hbbc{{\sc{H}}$\beta_{\rm BC}$\/}
\def\hbnc{{\sc{H}}$\beta_{\rm NC}$\/}
\def\mgii{{Mg\sc{ii}}$\lambda$2800\/}
\def\niv{{\sc{Niv]}}$\lambda$1486\/}
\def\nv{{\sc{Nv}}$\lambda$1240\/}
\def\ciii{{\sc{Ciii]}}$\lambda$1909\/}

\def\o4363{{\sc{[Oiii]}}$\lambda$4363\/}
\def\ovi{{\sc{Ovi}}$\lambda$1034\/}

\def\oiiiuv{{\sc{Oiii]}}$\lambda$1663\/}
\def\siiii{\ion{Si}{3}]$\lambda$1892\/}
\def\siv{\ion{Si}{4}$\lambda$1398\/}
\def\heiiuv{\ion{He}{2}$\lambda$1640\/}

\def\feiiuv{{Fe\sc{ii}}$_{\rm UV}$\/}

\def\fe{{\sc{Fe}}\/}

\def\fe76087{{\sc [Fe vii]}$\lambda$6087\/}
\def\oiii{{\sc [Oiii]}$\lambda$5007}

\def\kms{km~s$^{-1}$}
\def\ergss{ergs s$^{-1}$\/}
\def\aliii{\ion{Al}{3}$\lambda$1860\/}
\def\siii{\ion{Si}{2}]$\lambda$1306\/}
\def\gs{$\Gamma_{\rm S}$\/}

\received{......}
\begin{document}
\title{Average UV Quasar Spectra in the Context of Eigenvector 1:
A Baldwin Effect Governed by Eddington Ratio?}

\slugcomment{ }

\shorttitle{Average UV Quasar Spectra}

\shortauthors{Bachev et al. }
\author{
R. Bachev\altaffilmark{1,2}, P. Marziani\altaffilmark{3}, J. W. Sulentic\altaffilmark{1},
R. Zamanov\altaffilmark{3,4}, M. Calvani\altaffilmark{3}, D. Dultzin-Hacyan\altaffilmark{5}\\}

\altaffiltext{1}{Department of Physics and Astronomy, University of
Alabama, Tuscaloosa, AL 35487, USA; giacomo@merlot.astr.ua.edu}

\altaffiltext{2}{Institute of Astronomy, Sofia 1784, Bulgaria;
bachevr@astro.bas.bg}

\altaffiltext{3}{INAF, Osservatorio Astronomico di Padova, Vicolo
dell'Osservatorio 5,   I-35122 Padova, Italy; marziani@pd.astro.it;
calvani@pd.astro.it}

\altaffiltext{4}{National Astronomical Observatory Rozhen, POB 136, Smoljan,
BG-4700, Bulgaria}

\altaffiltext{5}{Instituto de Astronom\'\i a, UNAM, Apdo. Postal 70-264,
04510 Mexico D.F., Mexico; deborah@astroscu.unam.mx}

\begin{abstract}

We present composite UV spectra for low redshift Type 1 AGN binned
to exploit the information content of the Eigenvector 1 (E1)
parameter space. Composite spectra show high enough S/N and
spectral resolution to permit a decomposition of the \civ\ line
profile -- one of the strongest high-ionization lines (HIL) and
fundamental in defining E1 space. The simplest \civ\ decomposition
into narrow (NLR), broad (BLR) and very broad (VBLR) components
suggests that different components have an analog in \hb\ with two
major exceptions. VBLR emission is seen {\it only} in population
B (FWHM(\hbbc)$>$4000 \kms) sources. A blue shifted/asymmetric
BLR component is seen {\it only} in population A
(FWHM(\hbbc)$\leq$4000\kms) HIL such as \civ.  The blueshifted
component is thought to arise in a high ionization wind or
outflow. Our analysis suggests that such a wind can only be
produced in population A (almost all radio-quiet) sources where
the accretion rate is relatively high. We propose a model to
account for several differences between low- and high-ionization
line profiles. Part of the broad line emission is attributed to a
self-gravitating/fragmented region in an accretion disk. An inner
optically thick geometrically thin region of the flow may give
rise to a wind/outflow and produce the blueshifted HIL spectrum in
Pop. A sources.  The fragmented region may produce all or most of
the broad line emission in population B, which contains radio-quiet
and the majority of radio-loud quasars. Comparison between  {\em
broad} UV lines in radio-loud (RL) and radio-quiet (RQ) sources in
a single well populated E1 parameter space bin (B1) shows few
significant differences. Clear evidence is found for a significant
NLR CIV component in most radio-loud  sources. The BLR/VBLR
similarity in bin B1 provides circumstantial evidence in favor of
black hole (BH) spin, rather than BH mass or accretion rate, as a key
trigger in determining whether an object will be RL or RQ. We find a
ten-fold decrease in EW \civ\ with Eddington ratio (decreasing
from $\approx$1 to $\sim$0.01) while \nv\ shows no change. These
trends suggest a luminosity-independent "Baldwin effect" where
the physical driver may be the Eddington ratio.

\end{abstract}


\section{Introduction}

Understanding the geometry and kinematics of the central regions
in AGN is one of the major goals of twenty-first century
astronomy. Yet little is known beyond the necessity of invoking
accretion onto a supermassive object in order to account for the
enormous source luminosity. Given our inability to resolve the
central regions of even the nearest AGN we are left with three
approaches towards achieving our goal: (1) ideology:
sophisticated models driven entirely by
theories describing black hole and associated accretion disk
properties (e.g., Janiuk, {\. Z}ycki, \& Czerny 2000; Gammie,
Shapiro, \& McKinney 2004), (2) analogy: comparison with
better-resolved galactic objects (e.g. interacting binary stars;
Zamanov \& Marziani 2002; Mirabel 2004; Maccarone, Gallo, \&
Fender 2003) and (3) empiricism: detailed studies of phenomenology
in large and diverse samples of AGN (see Sulentic, Marziani \&
Dultzin-Hacyan 2000a for references until late 1999; Marziani et
al. 2003a, hereafter M03; Dietrich et al. 2002; Kuraszkiewicz et
al. 2004 for examples of more recent works). The last approach
fell on hard times until the end of the twentieth century in part
due to: (i) the apparent intractability of AGN phenomenology, (ii)
the failure of the best nebular physics models to reproduce the
observed spectra, and (iii) the low quality of so much of the
spectroscopic data. Thus we now know more about the morphology of
the extended radio and optical emission around quasars than the
underlying geometry/physics that drives them. One positive
development involves the application of reverberation mapping
towards estimation of the broad line region size (see Peterson
1993 for a thorough review; Horne et al., 2004 for up-to-date
expectations). The relevant questions are how
to interpret these data and, more importantly, how to select a
representative sample of sources for a reverberation mapping
study. Another positive development involves growing databases of
AGN spectra (most notably from the Sloan Digital Sky Survey
(SDSS); Richards et al. 2002; Schneider et al. 2002). Although
great part of the individual data still suffers from S/N limitations,
average or composite spectra of many objects can significantly
improve the situation (Vanden Berk et al. 2001; Sulentic et al.
2002). One of the most important questions is {\it how} to bin the
data. Clearly unrestricted composites have little value because
they hide the intrinsic dispersion in key parameters which
provides clues to the underlying physics (e.g., Francis et al.
1991). In a previous paper (Sulentic et al. 2002: hereafter Paper
I)  we argue that composite spectra binned within the Eigenvector
1 parameter space provide the most physically useful results.
Paper I presented composites of the H$\beta$\ spectral region and
this paper presents complementary composites for the UV spectral
region and especially for \civ.

We have identified a four dimensional parameter space that overcomes some 
of the problems mentioned above and that offers a new perspective on AGN
phenomenology (Sulentic et al. 2000a,b). The Eigenvector 1 (E1)
parameter space provides optimal discrimination/unification of
broad-line AGN classes (Boroson \& Green 1992; Marziani et al.
2001, hereafter M01; Boroson 2002). The adopted name reflects
recognition of the first two parameters in the pioneering work of
Boroson \& Green (1992). E1 as we now define it involves measures
of:

\begin{enumerate}

\item full width at half maximum of low ionization (LIL) broad
emission lines [FWHM(\hbbc)],

\item equivalent width ratio of optical FeII and \hbbc:
\rfe=EW(\feiiq\ blend)/EW(\hbbc),

\item soft X-ray photon index ($\Gamma_{\rm soft}$),

\item profile centroid velocity displacement (at FWHM) for \civ.

\end{enumerate}

In simplest terms these parameters can be said to measure
respectively: (1) velocity dispersion of LIL emitting gas, (2)
relative strength of LIL features thought to arise in the same
region, (3) strength of a soft X-ray component and (4) amplitude
of systematic motions in the high-ionization line (HIL) emitting
gas. In less simple (i.e. more model dependent) terms we likely
have: (i) two variables mainly sensitive to accretion rate
(FWHM(\hbbc) and \gs: Nicastro 2000; M01; Pounds, Done \& Osborne
1995; Czerny et al. 2004), (ii) three variables sensitive to
source orientation (all with the possible exclusion of \gs; e.g.,
Jackson \& Browne 1991; Wills et al. 1995; Marziani et al. 1996;
M01; Rokaki et al. 2003; Sulentic et al. 2003). \rfe\ is sensitive
to nebular physics (electron density $n_{\rm e}$\ and column
density $N_{\rm c}$).

An important addition that reflects the striking changes in
emission line properties across E1 involves introduction of the
population A and B concept. At present we arbitrarily define Pop.
A and B quasars as those with FWHM \hbbc\ less than or greater
than 4000 \kms. It is unclear whether the A-B concept reflects two
distinct quasar populations or simply the difference in mean
properties of sources at opposite ends of the E1 "main sequence".
The "elbow" shape of the sequence, and optical/X-ray indications
(Sulentic et al. 2000a,b; Komossa \& Meerschweinchen 2000) for a
"zone of avoidance" near FWHM(\hbbc) $\approx$ 4000 \kms, provide
some evidence in support of the former interpretation. If the
distinction is real then "A" is a largely RQ population
(including narrow line Seyfert 1 sources as an extremum), while
"B" contains a mix of broader line RQ plus most RL sources. The
population A-B distinction will prove useful in this study and
\civ\ will add support for the concept (\S 3.2 and \S 4.3).

If the E1 concept lives up to initial promise, it may be the strongest 
and most encompassing type-1 AGN unification yet discovered. This
possibility warrants strong tests of its strength and generality.
So far most of our effort has centered on the "optical plane" of
E1 (M01; Zamanov et al. 2002; Boroson 2002; Marziani et al.
2003a,b; Sulentic et al. 2003). This focus has been dictated by
the available X-ray and UV archival data that provide the other E1
parameters.  In the next section we present composite spectra for
\civ\ (and other UV lines) in the E1 context which complements
work already done for \hb.

\section{Generating Composite UV Spectra}

A total of about 700 reasonable quality (S/N\footnote{Here and
below S/N is the signal-to-noise ratio per resolution element,
measured in the continuum, blueward of \civ} $\geq$ 4--5) HST
spectra for n = 141 different AGN (quasars or bright Seyfert 1's;
n=61 are radio-loud with $\rm R_K \ga 100$, where $\rm R_K$\ is
the specific flux ratio between 4400 \AA\ and 6 cm) have been
extracted from the Hubble archive. The majority of the spectra (83
\%) were obtained with the FOS camera and the rest with STIS. The
FOS/BL data (40 \%) were recalibrated with the latest IRAF package
-- STPOA, which takes into account some systematic errors in the
wavelength calibration. No such recalibration is available for the
FOS/RD, where similar errors are known to exist, which ultimately
limits the wavelength accuracy of our composites to about 0.5--1.0
\AA\ (see for instance Kerber \& Rosa, 2000, for details on FOS
calibration).

Composite spectra cover the entire rest frame region between 1000--3000
\AA\ while individual source spectra will often cover only a part of it.
We applied the following procedures to compute the composite spectra:

\begin{enumerate}

\item Spectra for a specific source covering the same spectral region
were averaged together (if they were of comparable S/N).

\item A single spectrum for each object was created by adding the parts
covering different wavelength regions, with slight adjustments of the
continuum levels if necessary.

\item All spectra were corrected for Galactic extinction
following Schlegel et al. (1998). The typical extinction correction
$A_{\rm V}$ was about 0.1 -- 0.2 magnitude. No correction for the
intrinsic extinction was attempted.

\item The spectra were de-redshifted using the most accurate
available measurements (M03) or by using the latest accurate
values available in the literature whenever a source was not part
of the M03 sample\footnote{M03 did not publish most of their
refined redshift measures based on [OIII]$\lambda$5007 or the
narrow peak of H$\beta$ (see Zamanov et al. 2002); they are
available from the authors on request.}. It should be noted that
redshifts from V\'eron-Cetty \&  V\'eron (2001); hereafter VCV01,
sometimes differ considerably from our best values ($\Delta z$\
could be as large as 0.02).

\item Spectra were normalized by their own continuum flux in the wavelength
range 1430--1470 \AA, where no strong lines are observed/expected.

\item Composite spectra were generated by median-combining the averaged spectra of
individual AGN. This procedure ensures that only the most typical features will
be revealed, and that some unusual features like narrow absorption (often lost in
the noise), unusually strong (narrow) lines, noise spikes, etc. will be
omitted in the composites. A median also works better than an average when
the individual spectra do not cover the same spectral region (see \S 4 for
details regarding line profiles of the composites).
\end{enumerate}

The composites have been binned to 0.5 \AA\ per resolution element and have a
typical S/N$\sim$50 (see Table 1). The composite spectra are presented in Figure 1
with an enlargement of the 1100--2000 \AA\ region of greatest interest shown
in Figure 2. The latter show details of the strongest lines such as Ly\a, \civ, and
\ciii. The lines that we identified and measured (\S 3.1) are labeled.

We defined spectral types on the basis of source parameter
properties in the optical plane of E1 (Paper I). This  approach
differs from the one employed in analysis of the SDSS quasar
sample (Richards et al. 2002) where sources were binned {\it a
posteriori} in terms of the velocity shift between \civ\ and
\mgii. We are able to create useful composites for five E1 bins
with fixed FWHM(\hbbc) and \rfe\ ranges, exactly as done in Paper
I. The values of FWHM(\hbbc) and \rfe\ that provide an optical bin
assignment were taken from M03. In order to maximize the number of
sources in each bin we also include additional spectra for 
sources ($\approx$ 20\%\ of the total), not part of the M03 sample. 
This supplemental data involves:  1) sources with reasonable 
quality archival \civ\ data and 2) \hb\ spectra from literature 
sources with high enough S/N to permit assignment into one of 
the predefined E1 bins. Details on bin assignation for 
individual sources that are not part of the M03
sample will be provided in a future work.

Table 1 provides parameter measures and statistical properties for the
spectral bins. The format is: Column 1 -- spectral type designation
following Paper I; Col. 2 -- range of FWHM(\hbbc); Col. 3 -- range of \rfe;
Col. 4 -- number of sources used for the composite spectrum (number of RL source
in parenthesis). Each bin typically involves 15 -- 30 sources resulting in a S/N
in the continuum typically of 40 -- 60 (Col. 5).
Col. 6 and 7 present the average absolute $V$-magnitude (computed from
VCV01; $H_{\rm 0}=50$ \kms\ Mpc$^{-1}$; $q_{\rm 0}=0$ is adopted) and redshift
for sources used in the bin composite followed by the sample standard deviations.
We also present the average equivalent widths (EW) of \oiii\ and \hb\ for the
objects of the sample -- Col. 8 and 9 respectively.

There is a clear trend between  $\rm <M_V>$ and $z$, in the sense
that Pop. A sources show lower mean redshift and luminosity. This
is a known bias in the M03 dataset. M03 is a step towards our goal
to develop a sample complete to m$_V$=16.5. At this time the
radio-loud part of our sample is apparently complete (the standard
$V/V_{\rm max}$\ completeness test (e.g., Peterson 1997) gives
$\sim$0.8 compared to 0.3 for RQ sources). The radio loud sources
tend to be more luminous (due in part to relativistic beaming)
resulting in selecting more high redshift sources. The E1 optical
parameters {\it show no dependence on luminosity} to at least $z
\la 1$ and $\rm m_V \la$ 16.5 (Sulentic et al. 2004). We will
consider later possible UV line vs. source luminosity correlations
(see \S 3.3 and \S 4.3).

We also computed the sample standard deviation for the spectra in each bin.
Deviations from the composites at 1$\sigma$ confidence level are found to be
about 20\%\ and show no specific features after being normalized by the
composites themselves.

\section{UV Lines in the E1 Context}

\subsection{\feiiuv\ subtraction and line measurements}

Table 2 presents rest-frame EW and FWHM measures for UV lines in
the composite spectra. E1 bin designations are listed at the top
of the Table. Uncertainty estimates are given in parenthesis
following each measure. EW uncertainties are about 10--20\%\ for
the strongest lines (W$\ga$10 \AA) and approach 50\%\ for the
weaker lines. FWHM uncertainties are generally below 20--30\% with
the level of uncertainty depending on S/N of the composites as
well as line blending (e.g. \siv) in some cases. All measurements
were carried out after \feiiuv\ and continuum subtraction except
that these corrections were not applied to \mgii.

The \feiiuv\ template covering the 1000 -- 2000 \AA\ region (where
the most interesting lines are) was created with spectra for I Zw 1 
(see Marziani et al. 1996, M03 for details). The template was intensity
scaled and broadened in order to best match the observed FeII
pattern in the composite spectra. The results are summarized in
the last row of Table 2.  Figure 3 illustrates the deblending
results for two particularly important groups of lines: \lya\ +
\nv\ and \aliii\ + \siiii\ + \ciii. The first blend is important
because \nv\ is the only high-ionization line that shows no
significant Baldwin effect (Sulentic et al. 2000a, see also \S
4.5). The importance of the second blend involves the fact that
the intensity ratio \siiii / \ciii\ may be sensitive to density
(n$_e$) and/or may be affected by metallicity trends along E1
(M01; \S 4.3). We carried out a deblending procedure by assuming a
Lorentzian profile for both lines in the blend and allowing for
different component intensities and FWHM. A possible contamination 
by FeIII blends may contribute to some extend to the relatively 
poor fit of these lines in A3 bin.

We consider several interesting results that can be summarized as follows:

\begin{itemize}
\item  \ovi, \lya, \civ, \ciii\ and \mgii\ show a systematic
increase in equivalent width as one proceeds from extreme
population A (bin A3) towards the B bins. The most dramatic change
is actually seen in \mgii, \civ\, and also \ovi, where bin
averaged equivalent widths increase by almost a factor of four (a
factor $\sim$10 is seen in \civ\ if individual sources are
considered). While we adopted CIV centroid shift at FWHM as an E1
parameter other CIV measures follow the E1 sequence as well.

\item Contrary to the above lines \nv\ shows almost constant equivalent
width ($\approx$20 -- 25 \AA) across the E1 sequence. Other lines show
no clear trends within measurement uncertainties.

\item The \siiii / \ciii\ intensity ratio decreases monotonically
by a factor of four as one proceeds from bin A3 to B1$^+$ (from
0.8 to 0.2). See \S 4 for possible explanations.

\item We find a clear positive trend between FWHM and ionization
potential ($\chi$) for the principal broad lines. The correlation
coefficients are in the range 0.6--0.8. FWHM typically scales
linearly with $\chi$ up to $\chi \la 50$ eV and saturates beyond
that value. The scatter in this relation appears to be smaller for
A-bins. B-bins suggest that some LIL such as \mgii\ and \lya\
deviate from the relationship with larger than expected values of
FWHM. The linear part of the relation can be roughly described  by
the following expression: $FWHM=V_0 + \alpha\chi$, where $\alpha
\sim$ 60 -- 80 and V$_0$ from $~$1000 to $~$3000 \kms\ from bin A3
to B1$^+$. The saturation occurs at about 4000 \kms\ for A-bins
and at about 5000 -- 6000 \kms\ for B-bins.

\end{itemize}

\subsection{\civ\ Profile Decomposition}

The \civ\ line deserves special attention as the best studied, and
easiest to study, HIL. The \civ\ intensity and relative lack of
contamination by satellite lines, motivated its selection as the
representative HIL in the E1 parameter space. Profile centroid
shift at FWHM intensity was chosen as one of the most important E1
parameters (Sulentic et al. 2000b). The blueshifted component of
the \civ\ line seems to disappear in sources with FWHM \hbbc\
$\geq$4000 \kms\ providing further support for the pop. A-B
distinction. This fact and other previous works make it clear that
both HIL and LIL cannot be treated as single component lines. A
large part of the confusion in past broad line studies stems from
this fact. The simplest possible decomposition of our composite
\civ\ profiles was attempted after the following processing steps.
1) The continuum was subtracted by fitting a low-order Lagrange
polynomial over the region 1000--2000\AA.  2) A broadened and
scaled \feiiuv\ template was subtracted (\S 3.1, Table 2).
\feiiuv\ emission is not strong in any of the bin composites
however a reasonable correspondence between the optical and UV
FeII emission strengths can be traced in the sense that bin A3
shows the strongest \feiiuv\ emission. 3) Contaminating lines near
\civ: \niv\ and blended  \heiiuv\ + \oiiiuv\ were fitted with
Lorentzians and subtracted from the \civ\ wings. 4)  A narrow
component was subtracted from the \civ\ profile (see below).


Results of the \civ\ decomposition are given in Figure 4 and Table
3. Results from modelling our \hbbc\ composites  (Paper I) are
useful here.  We identified three reasonably unambiguous \hb\
components in the context of E1 and the Population A-B hypothesis.
These identifications, and the adopted profile fits, are largely
empirical at this stage. We focus on bins A2/A3 and B1 as
representative of the two AGN populations or E1 parameter space
extrema. Bin A1 is likely a mix of both population A and B sources
(or a transition between the two).

\begin{itemize}

\item \textbf {i) NLR}

\textbf{\hb:} We often see an inflection between the BLR and NLR
components of LIL \hb\ making profile decomposition less ambiguous
(Marziani et al, 1996). Bins A2/A3 show evidence for weak \hbnc\
emission ($EW\simeq3$ \AA) in residuals after modelling the broad
line components (Paper I). A more significant NLR component is
found in the majority of Pop B sources $<EW>\simeq5-6$ \AA\ with a
large scatter). If we assume a fixed ratio for \hbnc/\oiii, then
the Table 1 averages for EW(\oiii) imply a very weak \hbnc\
contribution in bins A2/A3. This ratio is about 0.11 for pop. B
sources which would imply EW(\hbnc) $\simeq1$ \AA\ for bins A2/A3
which is lower than what we actually subtracted except in the
"blue outlier" sources where \oiii\ is extremely weak and
blueshifted (Zamanov et al. 2002).

\textbf{\civ:} In our opinion it is clear that a narrow \civnc\
component exists in many, especially Pop. B, sources because we
can see a profile inflection (Sulentic \& Marziani 1999). The situation 
for \civ\ is more ambiguous than for \hbbc\ because of the lower
intrinsic resolution (and often S/N) of the \civ\ spectra which
less often show the inflection that can  guide a profile decomposition.
Either Gaussian or Lorentzian fits to the broad \civ\ profiles of
bin A1, B1 and B1$^+$\ composites show a narrow residual. Although
one cannot be sure that it is from an NLR component, that is the
simplest interpretation and is empirically justified by many
cases where NLR emission is unambiguously observed. New evidence
for an NLR component in, at least, pop B RL quasars is given in
\S 4.4. We therefore argue that an uncertain NLR subtraction is
better than no subtraction for the A1 and B-bins. No NLR component
was subtracted from bins A2/A3. The NLR profile width was assumed
FWHM$\la 2000$ \kms\ (taking into account the fact that \civ\ is
a doublet). Photoionization models suggest EW(\civnc)$\simeq$0.1
EW(\oiii) as a reasonable estimate (i.e. about 2 -- 4 \AA, using
the EW(\oiii) values from Table 1 and Paper I). 
We should point out that this EW ratio is rather similar to the flux 
ratio of the lines since the continuum shape does not seem to change 
significantly along the E1 bins (Fig. 1). Without inter-calibration 
of optical and UV fluxes, as well as knowledge
of the NLR extinction properties (e.g. Netzer \& Laor 1993), such
numbers are only indicative. See Sulentic \& Marziani (1999) for
more quantitative estimates of the effect of a \civ\ NLR
component on measures of FWHM and line shift. Our adopted
estimates of EW(\civnc)/EW(\civbc) show a large scatter (0.0 --
0.5) among individual sources in  A1 and B-bins.

\item \textbf{ii) BLR}

\textbf{\hb:} This is the defining broad-line component for Type 1
AGN. The Population A--B distinction is currently defined in terms
of sources with FWHM \hbbc\ smaller or larger than 4000 \kms.
Pop. A2/A3 bin composites are best fit by an unshifted Lorentzian
function (V\'eron-Cetty et al. 2001) while for Pop B bins a Gaussian
model works better. While the bin composite profiles are
unshifted, individual sources can show red or blue shifts of up to
4000 \kms. Profile shifts larger than about 600 \kms\ are seen
only in Pop. B sources (Sulentic et al. 2000a,b). Double peaked
profiles are also sometimes ($\leq$5\%) seen in bins B1$^+$ and
B1$^{++}$ \footnote{Bin B1$^{++}$\ was defined in Paper I for
sources with \rfe$<$0.5 and 8000 \kms\ $\la$ FWHM(\hbbc)$\la$12000
\kms}. Measures of FWHM \hbbc\ can be strongly influenced by the
NLR component (e.g. L/L$_{Edd}$\ and BH mass over- and
under-estimated, respectively).

\textbf{\civ:}  A strongly blueshifted line or line component is seen only
in Population A sources. The red half of the bin A2/A3 composite profiles can
be reasonably well fit with an unshifted Lorentzian model whose strength is
inferred from the amount of flux redward of the adopted source rest frame.
In this context extreme A3 sources should  have only a weak unshifted component
with the bulk of the line flux strongly blueshifted. We don't know if this
decomposition has any physical meaning, i.e. whether we have two distinct line
emitting regions or a single region, producing a complex profile. There are two
admittedly empirical reasons to favor the two-component model:

1. \hbbc\ is best fit with an unshifted Lorentzian and there is no {\it
a priori} physical reason why some HIL emission could not be
produced in the clouds responsible for the LIL emission. 
Current models do not rule out the possibility of significant 
CIV emission from the LIL clouds (Korista et al., 1997), although 
the accurate quantitative predictions require extensive knowledge 
of the physical conditions there. The derived
width of the best fit \civ\ Lorentzian components are in good
agreement with those estimated for \hbbc\ (about 2500 \kms, see
Paper I).

2. The fits in Figure 4 are consistent with a systematically
increasing  unshifted component as one proceeds from Bin A3 to
B1. The blueshifted (bulk of the emission in bins A2/A3) would be
strongly blue asymmetric and would truncate rather sharply near
the rest frame velocity, an interesting challenge for physical
models. This truncation is actually seen in sources like I Zw1 and
Ton 28 where a single component model is more appropriate
(Sulentic et al. 2000b, 2001). We cannot distinguish between a
Gaussian or Lorentzian model for the Pop. B BLR. The choice will
affect our estimate of the strength of the underlying VBLR.
Results from \hbbc\ fits lead us to favor, by analogy, a two
Gaussian model for \civ\ Pop. A sources involving unshifted BLR 
plus redshifted VBLR components. At this point we cannot place much
physical meaning on these  Lorentzian vs. Gaussian fits, they
simply reflect the relative sharpness of the line peak and the
extent of the profile wings.

\item \textbf{iii) VBLR}

\textbf{\hb:} There is no consensus whether or not the VBLR is a
distinct component (e.g. Corbin 1997; Snedden \& Gaskel, 2004;
Korista \& Goad 2004). Only  Pop B sources appear to show this "red 
shelf" or component that we model as a separate VBLR (see references 
in Sulentic et al. 2000a). We prefer to speak of the VBLR as a
distinct component for at least two reasons: 1) we often see an
inflection between the red wing of the best fit BLR Gaussian
and the red feature and 2) we have identified sources that only 
show a VBLR component (Sulentic et al. 2000c). Paper I composites 
suggest that the best current description of the components involve: 
i) a BLR with zero mean shift and FWHM $\sim 5000-6000$ \kms as well as 
2) a VBLR with mean redshift of about 5000\kms\ and FWHM$\simeq$10000 \kms.

\textbf{\civ:} No evidence for a significant VBLR component is
seen in population A bins paralleling the results for \hbbc.  A
double Gaussian decomposition is consistent with both BLR and VBLR
\civ\ components in Pop B bins\footnote{Although we fit A1 profile
with a single Lorentzian, we cannot rule out the possibility that
(some) A1 sources have both -- a weak VBLR and a weak blue-shifted
(wind) component, which along with a central unshifted component
mimic a single Lorentzian.}. FWHM of the components are in rough
agreement with those for \hb. Interestingly \civ\ VBLR line shifts 
appear to be lower ($\sim1000$ \kms) than those modeled in 
the \hbbc\ composites, however this may be a result of the fact that
the two-component fitting is not necessarily a unique operation.


Furthermore line shift values derived for such an ultra-broad component must 
be considered very uncertain especially given the low S/N of many HST spectra.

\end{itemize}

While we currently separate population A and B sources at
FWHM(\hbbc) = 4000 \kms, if the distinction has any meaning then
it will likely involve a complex combination of parameters. There
are two possible interpretations of the Pop. A-B concept: i) two
distinct AGN populations with different central source geometry and 
kinematics involving perhaps a critical Eddington ratio or ii)
a single "main sequence"
involving the majority of AGN and driven by a gradual change in
the ionization parameter/Eddington ratio (see M01). The "elbow"
shape of the source occupation in the optical plane of E1 is an
argument in favor of the former interpretation (i.e. it is not a
monotonic correlation). In any case the shape suggests that bin A1
will likely blend the parameter space properties of Pop. A and B
sources. Figure 4 and Tables 1, 2 support that view. A few sources
in bin A1 (e.g. NGC 1566 and Mrk 493) show very strong NLR
emission while others little or none. Our \civ\ profile analysis
supports the Population A-B idea in the sense that we find \civ\
line components unique to each (Pop. A -- blueshifted BLR, and Pop
B -- redshifted VBLR) but mutually exclusive.

Our results appear to be robust for sources with z$<$0.8. Richards
et al. (2002) created similar \civ\ composite spectra using part
of the much larger and higher mean redshift ($<z>\simeq1.8$) SDSS
quasar sample.  Their spectral resolution is similar to our own
but the S/N of individual spectra are usually much lower. They
binned spectra showing \civ\ into four groups (each involving
almost 200 sources) based on the centroid shift of \civ. We lack
LIL data that would allow us to assign unambiguous population A
and B tags to these bins but the two extreme groups (Richards et
al. composite D: most blueshifted and composite A: least shifted)
should, more or less, correspond to our A2/A3 and B1/B2 samples
respectively. If this statement is true then E1 predicts SDSS
composite A sample should contain many radio-loud sources and
composite D very few. We also note that SDSS composite A is more
sharply peaked than composite D, consistent with a stronger NLR
component. We applied our profile decomposition procedure to the
SDSS A and D composites kindly provided by G. Richards and the
results quantitatively verify the previous statement about their
similarity to  our lower redshift composites. The profiles are
also consistent with the presence of a growing (from D to A)
unshifted BLR component that further supports our two component
fits to the bin A2/A3 composites. Richards et al. (2002) interpret
the overall shift of \civ\ as the result of "missing" flux on the
red side of the profile. We think that the body of evidence,
within the E1 context, does not support this interpretation, e.g.
the blue wing of CIV$\lambda$1549 in extreme pop. A sources does
not resemble the blue wing of CIV$\lambda$1549 in the broader pop.
B sources. Similarly the blue wing on SDSS composite D is steeper
than that for composite A. If we relate missing flux to
obscuration then it would be surprising to find the strongest soft
X-ray emission from sources with the most extreme blueshifted
\civ\ profiles since both should hardly be related.

SDSS composites cannot be used to help us decide whether there is
a continuous "main sequence" in E1 or two distinct populations.
Caution is needed in interpreting the SDSS binning because we do
not know what the SDSS bin definition means in an E1 context,
especially when \mgii\ (wings) may be strongly affected by FeII 
emission. If \mgii\ follows the line shift behavior of \hbbc\ large 
relative shifts \civ\ -- \mgii\ can occur in pop. A because of \civ\
blueshifts and in pop. B because of possible \mgii\ redshifts. If 
\civ\ blueshifts (relative to rest frame and not \mgii) remain a
population A phenomenon at high redshift then the SDSS results
suggest that the fraction of population A sources is larger at
high redshift (75\% vs. our estimate that 60\% of quasars are
population A; Marziani et al. 2003b).  This would be consistent
with our suggestion that Pop. A sources represent the "seed"
population of young quasars in the Universe (Sulentic et al.
2000a; Mathur, 2000; Constantin \& Shields, 2003).

\subsection{RL -- RQ Differences and Luminosity Effects}

We can also compare RL vs. RQ properties using suitably
constructed composites. Selection of RQ and RL subsamples from a
single bin will ensure that we compare quasars with similar
physical and observational properties. An extreme comparison of
Pop. A vs. B will not be possible using a single bin. We would
have to compare an extreme pop. A bin with an extreme pop. B one.
Instead this will be a comparison between RL sources and RQ
sources with optical spectra indistinguishable from them. This
comparison offers hope to isolate subtle effects that might be
direct manifestations of radio loud activity. We must use a B bin
for this comparison because that is where most RL sources are
found. We choose B1 which is the most densely populated and where
RL and RQ objects are almost evenly represented. Radio (6cm) and
optical V-band fluxes were derived from data in the VCV01 catalog.
A radio-index (Kellermann et al. 1989) $R{\rm _K}$\ was computed
and all sources with log($ R{\rm _K}$)$>$2 are considered
radio-loud (see Sulentic et al. 2003). Figure 5 shows the RL and
RQ composites, as well as RL--RQ residuals, for both \lya\ and
\civ. We see evidence for differences: 1) most significantly,
stronger (EW$\sim$5\AA), NLR emission in the RL sources and 2) a
possible excess redshifted BLR (or VBLR) emission in RL sources
(this is also present in Figure 7 A--D composite difference of
Richards et al. 2002). The former is not a surprise (see e.g. Xu
et al. 1999) while the latter, if real, may be an important clue.
Since we know that some bin B1 RQ sources also show NLR emission
result 1) reflects a lower limit for the mean NLR contribution in
Bin B1 RL sources. A comparison shows also that the RL sources 
from this bin have on average stronger [OIII] and  \hbnc\ 
components (typically by about 30\%) than the RQ sources. 
There is no doubt that RQ sources exist with VBLR components 
as strong as the strongest observed among the radio-louds 
(e.g. Sulentic et al. 2000c) so further tests of the
red residual are needed.

In order to explore any luminosity correlation with composite
spectra we again restrict ourselves to bin B1. Sources were
separated into brighter and fainter groups using $ M{\rm _V} =
-25$ as the reference value. This value separates the bin B1
sample into two approximately even groups. Composite spectra
(covering 1000--2000\AA) and differences are shown in Fig. 6. One
can see a difference in the broad components of all the strongest
high ionization lines (\lya, \civ, \siv\ and \ciii) in the sense
that the equivalent width is larger for the lower luminosity
composite spectrum (the Baldwin effect? see \S 4.5 for a thorough
discussion). This difference was not seen in the RL vs. RQ
comparison. Fainter, lower ionization lines do not show any
difference, furthermore they are weak enough that any difference
would likely be undetectable.

\section{Discussion}

\subsection{The Relevance of the Median Composites}

The use of composite spectra raises an immediate question. Do
median composites yield typical quasar spectra for each bin? For
example, is a symmetric profile in a median composite the result
of equal numbers of blue and red asymmetric profiles? A visual
inspection of our sample spectra shows, e.g. for B-bins, about
60\% of sources with a more or less symmetric \civ\ profile, while
the remainder show 10, 20 and 10\%, respectively red, blue and
indeterminate (due to low S/N) asymmetries. This suggests that
median composites represent the most typical quasar spectrum and
that the changes between our spectral bins are real. If our bin
populations had $\ga$ 100 objects then we would use weighted
average composites, but medians appear to be a reasonable approach
at this time. It is also important to point out that bin sizes
were chosen to include sources that are statistically similar
within the average parameter measurement errors.

Another important question is whether or not an ensemble of line
profiles with a wide distribution of widths, strengths and asymmetries
will yield a representative profile of the same type  after combination.
We see that the many of the lines (especially the strongest) are better
fit by a Lorentzian whatever that means physically. Is it possible, for
instance, that an ensemble of Gaussian profiles will generate a
Lorentzian-like shape in composite spectra or the converse?  This might
lead to incorrect interpretation, and hence modeling, of line-forming
regions. In order to test this possibility we ran a simulation by combining
together about 100 artificial line profiles with a wide (normal)
distribution of widths and intensities, resembling the real situation to
some extent. We performed simulation with both Gaussian and Lorentzian profiles,
creating median and average composites from the individual
pseudo-spectra.  The results suggest that profile type is preserved,
i.e. a composite of Gaussians or Lorentzians will preserve their
respective profile forms in a composite spectrum.

\subsection{Inferences From CIV Profile Decomposition}

Proper decomposition of emission lines in composite spectra along
the E1 sequence offers insights into changes in BLR geometry and
kinematics. Several results from this and previous papers
provide interesting clues.

Asymmetric blueshifted HIL emission is seen almost exclusively in
population A sources. In the most extreme cases a blueshift is
also observed in the unusually weak \oiii\ features (Zamanov et
al. 2002; Marziani et al. 2003c). One simple interpretation that
immediately suggests itself involves some kind of outflow from an
optically thick emitting region into an optically thin NLR. In a
scenario where the broad lines arise from an accretion disk this
could involve a high ionization disk outflow or wind. It would be
strongest in population A sources and would gradually weaken or
even disappear as one proceeds along the E1 sequence toward
population B. Along the same sequence an unshifted/symmetric
Lorentzian HIL component (which may be analogous to the dominant
LIL component) may be growing, in exactly the opposite sense of
the blueshifted feature. In fact some theoretical models predict 
Lorentz-like profile wings (e.g. Penston et al., 1990; Dumont \& 
Collin-Souffrin 1990). The profile may become more Gaussian in
the population B region. We have argued in a pre-E1 context that
HIL vs. LIL line shifts appeared to be completely uncorrelated for
RQ sources and marginally correlated for the RL minority (Marziani
et al. 1996). Composite spectra, binned in an E1 context, suggest
that CIV and \hbbc\ may show analogous emission components except
for the blueshifted HIL feature seen only in pop. A that are
almost always RQ AGN.

A very broad and redshifted VBLR component is only seen in
population B sources and may be present in both HIL and LIL.
Since NLR-BLR-VBLR inflections are rarely seen, there may be less
spatial/kinematic decoupling between these regions for CIV than is 
seen for \hbbc, where clear inflections are common. That is far from
proven because we also see LIL profiles with little or no
inflection between components. Inflections may be driven by
kinematic and/or orientation. There are some indications that the
region producing \civnc\ emission is closer to the center that the
region producing \hbnc\ (Sulentic \& Marziani 1999). The \civ\
VBLR may be most simply interpreted as arising at the inner edge
of the classical BLR in analogy to \hbbc\ VBLR (Marziani et al.
1993; Sulentic et al. 2000c). The lower redshift of the \civ\ Pop.
B. sources VBLR may also indicate that a high ionization wind is
still present in the VLBR since the \civ\ component is still
blueshifted with respect to the corresponding \hb\ component.

\subsection{BLR Structure and Kinematics}

So far there is no single picture capable of explaining the overall
properties of the broad line emitting region. Currently fashionable models
involving line emitting clouds near an optically thick accretion disk suffer a
number of difficulties. If the motions of the clouds are $Keplerian$ then they
will collide with the disk and will likely be destroyed on short time
scales. If one adopts an outflow/inflow model then the entire profile
will show large red or blue-shifts because, the standard optically
thick disk will obscure emission from the far side of a flow, while
in reality large shifts in line profiles are rare.

Alternatively, emission lines originating from a geometrically thin,
optically thick disk (e.g. Dumont \& Collin-Souffrin 1990) will show
extremely small FWHM when observed face-on. This situation may be relevant
for pop. A sources but latest work suggests that the situation is different
for RL and, by analogy, other pop. B sources. The most jet-aligned
superluminal RL sources, which, should have negligible disk rotation contribution
to LIL FWHM, show FWHM$\simeq$3000--4000 \kms\ (Rokaki et al., 2003;
Sulentic et al. 2003).  It is however true that core-dominated (CD) and
superluminal sources with $i \rightarrow 0^\circ$ show systematically narrower
profiles than other (FRII) RL AGN (Sulentic et al. 2003). CD sources are
displaced towards lower FWHM(\hbbc) in the optical E1 plane, which is 
consistent with orientation predictions. However, even these near
pole-on sources lie well above many RQ sources with similar \rfe, i.e.,
they have larger FWHM(\hbbc) for the same \rfe\ (Sulentic et al. 2003).
The results for radio loud sources suggest that we need a considerable velocity
dispersion in the vertical direction to account for observed LIL widths
(FWHM(\hbbc)$\approx 3000$ \kms).

The above results might be telling us that the bulk of line emission in RL and
Population B RQ sources does not arise from a standard optically thick
accretion disk especially when we observe FWHM LIL/HIL $\sim$ 3--4000 \kms\
in sources reasonably interpreted as almost disk face-on. One candidate for the
line emitting region involves the outer self-gravitating part of the
disk (Collin \& Hur\'{e}, 2001; Bian \& Zhao, 2002). The disk is expected to
become gravitationally unstable at some radius \rsg, where it will break up
into rings or discrete self-gravitating clouds. The vertical velocity dispersion of
such clouds may be amplified by cloud-cloud encounters. Some studies suggest that
the disk can become self-gravitating at a distance from the
central black hole as small as 100--200 \rg, where \rg\ is the
Schwarzschild radius (see for instance Collin \& Hur\'{e}, 2001).
A detailed theory for a self-gravitating disks is not yet available,
however it is expected that the radius of the self-gravitating region
will depend on the central black hole mass and Eddington ratio (and two
related and very important parameters: viscosity, $\alpha$, and disk
opacity). There is no consensus on the correct dependence of \rsg\ on
parameters like $M_{\rm BH}$ or Eddington ratio
$\dot m=L/L_{\rm Edd}$ (here we assume that the accretion efficiency ($\eta$)
does not change with luminosity, otherwise $\dot m=\dot M/\dot M_{\rm Edd}$,
where $\dot M_{\rm Edd}=L_{\rm Edd}/(\eta c^{2})$.

A very simple approach for exploring this idea involves
application of the Toomre stability criterion to a standard
Shakura--Sunyaev disk (Shakura \& Sunyaev 1973). This will give an
estimate of the radius where the stability parameter (Q) becomes
approximately 1.  Different models of the self-graviting region
often predict different \rsg\ dependence on the accretion
parameters because of  different initial assumptions about the
outer regions of the disk (i.e. opacity, pressure, etc). Bian \&
Zhao (2003) obtain in a recent work:  \rsg $\propto\dot m^{-0.49}
M_{\rm BH}^{-1.16}$ based on standard accretion disk solutions
assuming dominant Kramers opacity. They assume that it is equal to
the radius of the BLR $R_{\rm BLR}$\ (here and below \rsg\ is
expressed in \rg). Collin \& Hur\'{e} (2001) adopt a numerical
approach with realistic opacities and get a complex \rsg\
dependence roughly approximated by \rsg$\propto\dot m^{0.16}
M_{\rm BH}^{-0.46}$ for the region log($M_{\rm BH}$)=7 -- 9,
\.{m}$=0.01 - 1$. Following these authors, for a BH mass M=10$^7$
\msol\ and Eddington ratio $\simeq$1, \rsg$\approx 10000$ \rg,
while for M=10$^9$ \msol\ and \.{m}$\simeq$0.01, \rsg\ is less
than about 500 \rg.

On the other hand we can consider "superluminal" radio-loud
sources with the narrowest RL line profiles: FWHM(\hbbc)$\la$ 3000
\kms, and the narrowest Pop. A sources, with FWHM(\hbbc)$\la$ 1000
\kms, also likely to be observed at $i\approx0^\circ$. Assuming
orbital motion with Keplerian angular velocity ($\Omega_{\rm K}$),
one can write for these apparently face-on sources:

\begin{equation} FWHM \simeq \Delta v \simeq \nu \Omega_{\rm K} R_{\rm SG} \end{equation}

where $\Delta v$ is the vertical velocity dispersion, assumed to
be proportional to the $Keplerian$ velocity by a factor of $\nu$.
A reasonable guess for $\nu$ is about 0.1--0.2. One can easily
show that FWHM of ~3000 \kms\ implies a radius of about 5000 \rg\
while 1000 \kms\ implies ~500 \rg, values quite close to the size
of the self-graviting region assumed above. Even if the agreement
is perhaps fortuitous, and the exact \rsg\ dependence is far from
certain, it is reasonable to conclude that \rsg\ ($\sim$ $R_{\rm
BLR}$) may be smaller in Pop B sources by a factor $\sim 10$, and
that this may leave a very "small" emitting surface for any
standard optically-thick geometrically-thin disk component. Part
of the line profile may be produced in the disk if it is
illuminated by a geometrically thick inner region or by disk
warping (e.g. Bachev, 1999) and if it is not extremely hot (which
would completely ionize the lower-ionization species). On the
other hand, the presence of a blue-shifted component can be most
simply explained as the signature of an outflow from an optically
thick disk. The disk-wind scenario can produce blue-shifted
high-ionization lines as already suggested in several works (e. g.
Murray \& Chiang 1997; Bottorff et al. 1997). If our
interpretation is correct this implies that the wind strength
increases toward the most extreme population A sources (bins A2/A3
with strongest \rfe). Population B sources may have a much lower,
or may not develop, a wind component. Following these
considerations we propose a simple BLR scenario (see also  Fig. 1
from Collin \& Hur\'{e}, 2001):

\begin{itemize}

\item A "hot" innermost accretion region, located inside some
transition radius $R_{\rm Tr} \la 50-100$\rg). Our data leave the
structure of this region open to speculation, although we may
reasonably assume that the continuum region (or part of a
geometrically thick, optically thin advection-dominated accretion
flow (ADAF; see Narayan, Mahadevan \& Quataert 1998 for an
exhaustive review) can heat the outer line-emitting region.

\item An optically thick, geometrically thin disk whose extent
decreases by a factor $\sim$10 from Pop. A to Pop B. The heated
disk produces a wind that is responsible for the blueshifted \civ\
component. The outer region of the disk may emit \hbbc\ and other
LILs as well in Pop. A sources. The disk may become almost
negligible in Pop B sources because \rsg\ may become as small as
100 -- 200 \rg\ (e.g. Bachev \& Strigachev, 2004).

\item A self-gravitating and fragmented zone where line emission might arise in
dense gas clumps. This region may be the {\em entire BLR} for Pop. B sources
while it may contribute mainly to LIL emission in Pop. A objects.
\end{itemize}

Within the context of this model + E1 we can eventually explain several
observational results from this paper as well as from previous works:

\begin{itemize}

\item The different vertical velocity dispersion in the self-gravitating
zone for Pop. B (closer) and Pop. A (farther) sources accounts for
different line widths at $i\rightarrow0^\circ$;

\item After subtraction of an unshifted/symmetric Lorentzian component
the blueshifted residual shows reasonable persistency in shape -- the
intensity may change but not the overall shape. The blue component
becomes marginal at bin A1. This behavior is consistent with the idea
that the blue component comes from a disk-generated wind, where the
kinematics (velocity field) is strongly affected by the detailed
properties (e.g. magnetic field structure and terminal wind velocity).
The overall shape of a line arising in a disk wind will be less
dependent on the radial distance where the wind originates. In other
words, our model provides an easy explanation why population A sources
tend to show decoupled HIL--LIL components while Pop. B sources show
more similar \civ\ and \hbbc\ profiles. In the most extreme cases,
there are sources where the entire \civ\ profile resembles the blue
component alone (e.g. I Zw 1, Ton 28; Sulentic et al. 2000b).

Leighly (2004) presented an interesting explanation of the apparent
anticorrelation between the strengths of the central and the blue-shifted
(wind) components. The wind that gives rise to the blue-shifted component
might filter the EUV ionizing radiation that would otherwise reach the outer,
self-graviting parts (where the unshifted line component might originates).
The wind cannot however filter effectively the hard X-rays that are
reprocessed in LIL emission in the same region.

An alternative approach is to ascribe the entire profile of CIV in A2/A3 bins
to an outflow (wind). In such a case the redward emission is seen either
because the disk is (partially) transparent to CIV photons or because of
higher inclinations, resulting in significant projection of the Keplerian/radial
component of the wind velocity onto the line of sight. Both possibilities seem
somewhat unlikely, at least for these extreme bins.

\item There are trends between $\chi$\ and FWHM (e.g. Tytler \&
Fan 1992). Pop. A sources can be explained with a split BLR: inner
HIL and an outer LIL emitting regions. Since we expect that gas
motion will retain its circular velocity in the wind and will
acquire an upward and outward velocity component,  the velocity
dispersion in the wind will be higher than for a Keplerian
velocity component in the disk. LIL emission from the outer,
fragmented region will therefore be narrower. In Pop B sources we
expect a "normal" BLR stratification: HIL emitted closer to the
central continuum source than LIL. In general, we expect that both
HIL and LIL will be emitted rather close to the central continuum
source, accounting for the rather high ionization spectrum. This
outline requires testing with photoionization models.

\end{itemize}

Reverberation mapping indicates continuum-response times  of
\ciii\  consistent with the one of \hbbc\ (at least in NGC 5548;
Clavel et al. 1991). This result suggests that \ciii\ is primarily
produced in the low-ionization emitting part of the BLR. Under the
assumption of nearly constant [Si/C] abundance ratio, the behavior
of the \siiii/\ciii\ intensity ratio along the E1 sequence can then
be understood in terms of increasing electron density toward type
A3 (M01) in the LIL-emitting BLR. The critical \ne\ of the \ciii\
line is $\sim 10^{9.5}$ \cm3, much lower than that of the \siiii\
line, $\sim 10^{11}$\ \cm3. Therefore, for \ne $\ga 10^{10}$ \cm3,
the \ciii\ line should be more and more collisionally suppressed
with increasing \ne, while the \siiii\ line should become favored.
We also observe a consistent increase in the strength of the \aliii\
line, with the intensity ratio \aliii/\ciii\ reaching
$\sim 0.8$ for spectral type A3 (\ne $\sim 10^{11}$ \cm3). The \aliii\ line is
expected to strengthen with increasing electron density (Baldwin et
al. 1996; Korista et al. 1997). However, as pointed out by the referee,
chemical abundance may not be constant along the E1 sequence. We
do not have information on the [Si/C] and [Al/C] abundance ratios.
Intensity ratios involving nitrogen lines suggests that
super-solar metallicity is appropriate for extreme population A
(i.e., A3) sources. From Table 2, we have \nv/\civ $\approx1$ and
\nv/\heiiuv $\approx$3 for A3, while for A1 \nv/\civ $\approx0.25$, 
and \nv/\heiiuv $\approx$1.5. It is intriguing that we are
observing again, for spectral type A3, line ratios that are more
frequently found in high-$z$ quasars (Hamann \& Ferland 1999).
Highly supersolar nitrogen enrichment may have been produced by
star formation with a top-loaded initial mass function (Hamann \&
Ferland 1992; \S 4.5). It is possible that changes of line
ratios from Pop. B to Pop. A and along the A-spectral type are due
to {\em both} an \ne-increase and a metallicity increase.
Quantifying these effects is a challenge for further work.

\subsection{Radio Loud Sources: Black Hole Spin {\em and} Host Morphology?}

Comparison of RL and RQ line profiles (\S 3.2) in the E1 context
should begin with a comparison of population A and B sources. The
differences that we find between Pop. A and B HIL CIV composites
are even larger than for LIL \hbbc. We can say with some
confidence that sources showing low EW, blueshifted and
blue asymmetric \civ\ have low probability (P$<$1\%) to be
radio-loud. Pop.B sources have a reasonably high probability to be RL 
(P$\sim$50\%; Marziani et al. 2003b). Since not all pop. B sources are 
RL we focus (\S 3.3) on a particular bin (B1) within population B where we
can compare \civ\ properties for RQ and RL AGN that appear to be
spectroscopically indistinguishable. Our RL-RQ difference spectrum
is shown in Figure 5. We see evidence for a significantly stronger
NLR component in pop. B RL sources. This might be
explained if a significant part of the narrow line emission is
produced in interactions between a radio jet and the ambient
medium, forming presumably a narrow-line-bright cocoon around the
jet (e. g., Steffen et al. 1997; Axon et al. 1998). Aside from a
possible excess of redshifted broad line gas in the RL sources,
Pop. A RL and
RQ line profiles look very similar as was also found for \hbbc\
(Marziani et al. 2003b). The VBLR component, which might be
related to this redshifted gas, is seen in both RL and RQ Pop. B AGN
(Sulentic et al. 2000c) and, therefore, is not a RL signature. 
One is tempted to consider host galaxy
morphology as a major difference between the RL and RQ sources in
bin B1. RL sources tend to be hosted in ellipticals (Pagani et
al.  2003) while a typical pop. B RQ source involves a classical
Seyfert 1 like NGC 5548, hosted in a spiral. Perhaps the Pop. B
emission line spectrum indicates that RL activity is possible as
far as physical conditions near the BH are concerned but host
morphology dictates whether it happens or not. If it were that
simple however we might expect to see many frustrated RL sources
involving trapped core emission (unless it is self quenching under
that circumstance).

The bin B1 results emphasize the fact that Pop. B RQ and RL
sources are spectroscopically indistinguishable as far as {\em
broad} lines are concerned. This may be telling us that
radio-loudness is not directly dependent on, or strongly
influencing, BLR structure (and therefore -- the structure of
the accretion disk, at least at $r \ga 100$ \rg). This makes a
circumstantial case for the role of black hole spin, which
is unlikely to have any influence on the accretion disk structure
outside 10--100 \rg\ where the UV line emission likely originates, at
least if the Bardeen-Petterson effect (Bardeen \& Petterson, 1975)
is not in play. The black hole spin can, however, play a
significant role in jet production (e.g. Blandford \& Znajek,
1977). The fact that the majority of RL sources occupy B-bins may
be associated with a change of the innermost accretion disk
structure (from a thin disk to an ADAF?, R\`{o}zanska \& Czerny,
2000, and references therein) that is probably triggered by the
lower accretion rate (or by some function of the black hole mass
and the accretion rate). If true, the accretion disk structure
change (from thin disk to ADAF) will be a necessary but not
sufficient condition for radio-loudness (i.e. the ultimate trigger
will be the black hole spin, subject to the environmental caveat
mentioned above). In this scenario only B-bin objects which
contain a rapidly spinning black hole would be able to
produce powerful jets.

\subsection []{A Local "Baldwin Effect" Driven by the Eddington Ratio?\footnote{
We call the $L/L_{\rm Edd}$-dependent Baldwin effect also a "local" 
Baldwin effect since it is defined through low-z, low to moderate 
luminosity quasars and Seyfert 1 nuclei, in opposition to the 
canonical Baldwin effect which emerges considering high-z, 
high-luminosity quasars}}

Measurement of broad line EW's reveals both a trend and
the lack of a trend (see also \S 3.1):

\begin{itemize}
\item \lya, \civ\, \mgii\ and  \ovi\ show a systematic EW decrease
by factors of 2 -- 4 as one proceeds from  bin B1$^+$\ to A3. Table 2
shows the same trend for other lines but caution is needed when
interpreting blended lines and data for lines with EW$\la$5\AA\
that have estimated uncertainties $\ga$30\%\ at the 1-$\sigma$\
confidence level.

\item The \nv\ line shows no trend with EW(\nv) $\approx$
20--25\AA\ in all spectral bins.
\end{itemize}

These results essentially describe the so-called "Baldwin
effect" (hereafter BE; originally defined as an
anti-correlation between rest frame EW \civ\ and source
luminosity at 1450\AA: Baldwin 1977). It is well known that \nv\ does not
follow the BE trend (see Sulentic et al. 2000a for references
before 1999; Dietrich et al. 2002; Green et al. 2001; Croom et al.
2002). While the original BE showed a tight correlation involving a
few tens of sources, more recent work (Dietrich et al. 2002;
Warner et al. 2003) show it to be  a very loose correlation
that only becomes significant over a wide range in continuum
luminosity $\log(\lambda L_\lambda) $=42--48 \ergss\ (e.g. Kinney
et al. 1990). Considering the large dispersion it is not
surprising that several studies using small samples found no
evidence for a BE (see Sulentic et al. 2000a for numerical
simulations and a thorough discussion of the issue).

The BE-like trend found between our spectral bins does {\em not}
represent a dependence on luminosity. The E1 sequence is
independent of source luminosity (Sulentic et al. 2000b) and, is
more likely governed by the Eddington ratio  (M01; Marziani et al.
2003b). In fact Pop. B spectral bins contain more high luminosity
sources than Pop. A because RL in our sample are
overluminous, and overrepresented, relative to
RQ sources (see also Boroson \& Green 1992). Our BE
trend is also {\em not} BH mass-dependent (e.g., Warner et al.
2003): the largest BH masses are again found in population B
(M03). If we compare the average \.{m} values for our bins
(from M03b) with EW(\civ)
measures from our composite spectra we obtain the following
approximate relation: $\log$ EW(\civ) $\simeq 1.35 - 0.67 \log
\frac{L}{M}$ excluding bin B1$^+$. Our models predict that
B1$^+$ sources are simply more inclined B1 sources (M01). Including
B1$^+$ yields  $ \log$ EW(\civ)$ \simeq
3.22 - 0.39 \log \frac{L}{M}$ using $L/M$\ bin averages from M03b or $
\log$ EW(\civ)$ \simeq 3.54 - 0.47 \log \frac{L}{M}$\
if bin averages are taken from this paper (L/M is measured in Solar units).
f we consider all sources (RQ and RL) a robust fitting method
yields $ \log$ EW(\civ)$ \simeq 3.23 - 0.36 \log \frac{L}{M}$,
which becomes $ \log$ EW(\civ) $\simeq 3.40 - 0.41 \log \frac{L}{M}$
for RQ sources alone.

Parameter dispersion in the Baldwin effect (from EW(\civ)
$\approx$ 20 -- 200\AA) is well reproduced
by sources spanning the range $\sim$ 0.02 -- 1 in $\dot{m}$\
regardless of the fitting solution details. Highest $\dot{m}$
sources show the lowest EW(\civ) and it is a well-established
fact that NLSy1s show the lowest EW(\civ) (Marziani et al. 1996;
Rogriguez-Pascual et al. 1997; Sulentic et al. 2000a) and are
likely to radiate close to the Eddington limit. Many known NLSy1s
are relatively low redshift/luminosity sources that tend
to blur standard BE correlations at the low luminosity end (see
e.g., Brotherton \& Francis 1999) because they show EW(\civ)
similar to the high redshift quasars (Sulentic et al. 2000a).
Sulentic et al. (2000a) suggested that the BE might be caused by
preferential (probably driven by selection effects and/or intrinsic
evolution) detection of large $\dot{m}$ sources at high redshift.
In a flux limited sample, where $L$ and $z$\
are highly correlated, an "evolutionary" (redshift-dependent)
Baldwin effect is expected and is indeed found in the Large Bright
Quasar Survey (Green et al. 2001). Given a sample that is complete
in terms of $\dot{m}$, the BE may survive with even larger scatter or
disappear depending on the relative importance of evolutionary and
selection effects (see Dietrich et al. 2002). Deeper surveys may
detect larger EW(\civ) sources at high redshift: see for example
the distribution of EW(\civ) at $z\ga4$ in Constantin et al.
(2002). Intriguingly, they detect several sources with
EW(\civ)$\ga$50\AA\ at $\log \rm L \approx 46.5$ \ergss.

Interpretation of the BE has been highly controversial with
suggestions encompassing orientation effects (Netzer 1985),
selection effects (Sulentic et al. 2000a, and references therein),
black hole mass (Wandel 1999; Warner et al. 2003), Eddington ratio
(Sulentic et al. 2000a,  Shang et al. 2003; Baskin \& Laor 2004).
A continuum softening with increasing luminosity (Wandel 1999) is
a less model-dependent explanation. Warner et al. (2003) suggest
that the driving parameter may be the central black hole mass.
These authors computed the BH mass using a modified Kaspi et al.
(2000) relation, employing FWHM \civ\ instead of FWHM \hb\ and
assuming virial motions. This {\it does not demonstrate} that mass
is driving the BE because it is assumed to be $M\propto L^{0.7}$,
and this assumption makes the computation in part circular.
Furthermore, we question the validity of estimating BH masses
using \civ\ (e.g. Vestergaard 2002 and Warner et al. 2003; 2004).
While \hbbc\ shows average broad line profiles that can
be argued to arise in virialized clouds, \civ\ is blue asymmetric
and blue shifted in about 60\%\ of sources. In many cases the shift 
is significant and in some -- virtually all of the line flux is 
displaced to the blue side of the local rest frame. 
It is not at all clear that the clouds (or
wind?) producing the HIL emission can be assumed to be virialized.
Finally failure to correct for NLR emission in the CIV profile
will contribute to the dispersion in BH mass that might be present.
NLR emission tends to decrease the dispersion BLR FWHM measures
(FWHM measures for pop. B become more like Pop A).
Figure 5 shows that NLR uncorrected FWHM CIV measures will
underestimate BH mass and overestimate $L/L_{\rm Edd}$ values for
RL sources.

Our work shows that a \civnc\ exists in, at least, Pop. B sources
and that a significant part (if not all) of \civ\ emission
in bins A2/A3 arises in a wind. The terminal wind velocity 
could be only indirectly related to Keplerian (i.e. virialized)
motions. The overall velocity dispersion is even less likely to
be related to virial motion in near face-on sources, where the
accretion disk likely obscures the receding part of the flow. The
\civ\ line profile in such cases is almost fully blueshifted with
respect to the rest frame. While any correlation with mass will be
little affected over several orders of magnitude (Warner et al.
2003), any more fundamental correlation with $\dot{m}$ may be
entirely washed out ($\dot{m}$\ range is a factor $\la$50; Woo \&
Urry 2002). Comparing FWHM H$\beta$ and CIV in different spectral
bins (Pop. B with and without \civnc\ subtraction) suggests that
errors could be at least a factor $\sim$10 for some objects (since, 
for example, the FWHM ratio \civ\/\hb\ is $\approx$ 3 in IZw1; Marziani et
al. 1996; 2003b). However, a shallow trend between EW(\civ) and $L/L_{\rm
Edd}$\ is noted in a large sample of quasars of redshift $0\la z
\la 5$ (Warner et al. 2004), which seems to support our results for
$z\la $0.8.

Another circumstantial element supporting a role for the accretion
rate $\dot{m}$ in the BE involves the constancy of EW(\nv) and the
increasing intensity ratio \nv/\civ\  along the sequence from
B1$^+$ to A3. The large values of EW(\nv) for all spectral types
are suggestive of solar or supersolar metallicity (Turner et al.
2003; Bentz \& Osmer 2004). It is not unreasonable to suppose
that N may have been enhanced by vigorous formation of massive
stars that burn H via the CNO cycle in the AGN circumnuclear
regions. The largest enrichment may have occurred in relatively
young or rejuvenated quasars radiating at large accretion rates as
NLSy1s are thought to be (several NLSy1s hosts may be merging
dwarf or barred galaxies; Krongold et al. 2001; Crenshaw et al.
2003).

\section{Conclusions}

We computed and analyzed composite UV spectra for 5 different
spectral bins along the main sequence of the E1 optical parameter
plane. Data were analyzed following Paper I, which considered
similar composite spectra for the optical region involving \hb,
our representative LIL. We measured the FWHM and equivalent width
of a number of emission lines focussing on \civ, our
representative HIL. Our main results include:

\begin{itemize}

\item  We decompose HIL \civ\ into possible NLR, BLR and VBLR
components many of which have LIL analogs. We consider them to be
distinct line emitting regions. Type 2 AGN show only NLR LIL and
HIL emission. Type 1 AGN are know with pure BLR, pure VBLR and,
most often, composite BLR+VBLR (with and without NLR) spectra if
they belong to Population B. A strong asymmetric blueshifted BLR
HIL component is unique to population A (bins A2/A3) sources. A
reasonably strong and redshifted VBLR is unique to HIL and LIL in
population B (bins B1 and B1$^+$) sources.

\item We infer a \ne\ trend along the E1 sequence using the ratio
of the semi-forbidden lines \siiii\ and \ciii. Under the assumption 
of a nearly constant chemical abundance along E1, we find that
density increases by about an order of magnitude between bins 
B1$^+$\ and A3. However, there are indications that both -- density
and metallicity may increase from extreme B-bins to extreme A-bins.

\item We compare sub-samples of RL and RQ sources in bin B1. We
find an excess of NLR and redshifted broad line emission in the RL
subsample. The former represents virtual proof that  NLR line
emission is present in the spectra of many type 1 AGN.

\item We advance the following empirically driven model for the
structure of the accretion flow:

\begin{enumerate}

\item An inner hot optically thin (possibly ADAF) disk, located
inside $R_{\rm Tr}\la100$ \rg.

\item A standard thin disk, between $R_{\rm Tr}$ and $R_{\rm SG}$,
producing most of the optical continuum. This is the place where a
wind may originate and produce the blue-shifted \civ\ component.

\item An outer self-gravitating region at $r>R_{\rm SG}$, producing
most of the broad-line emission.

\end{enumerate}

$R_{\rm Tr}$ and $R_{\rm SG}$ are functions of the accretion
parameters. A n eventual discontinuity in their behavior as functions
of M and \.{m} may account for the existence of different quasar
populations (e.g. Pop. A, B).

\item We suggest that the Eddington ratio is the primary driver of
the Baldwin effect rather than source luminosity. This suggestion
is motivated by the robust BE that we find in E1 which is
luminosity independent. This explains why high redshift sources
might preferentially show  low EW and blueshifted \civ\ if Pop. A
sources are the "seed" (i.e. youngest high accreting lowest mass
BH) AGN. This interpretation is supported by comparison  of our
low redshift sample with first results from higher redshift SDSS
quasars.   The latter show an increase in the fraction of sources
with a blueshifted \civ\  HIL -- a defining characteristic of Pop.
A AGN.

\end{itemize}

\begin{acknowledgements}

RB and JWS  acknowledge the kind hospitality and the
financial support from the Osservatorio Astronomico di Padova
where much of this work has been done. We thank Dr. G. Richards
for providing his SDSS composite spectra in an electronic
form. We are indebted to an anonymous referee for his/her
constructive criticism.

\end{acknowledgements}

\clearpage

\begin{deluxetable}{ccccccccc}
\tabletypesize{\scriptsize}  \tablenum{1} \tablewidth{0pt} \tablecaption{{\sc
Definition of spectral types}} \tablehead{ \colhead{Bin} &\colhead{FWHM(\hbbc)}
& \colhead{\rfe} & \colhead{N$_{\rm tot}$ (N$_{RL}$)} & \colhead{S/N}&
\colhead{$<$M$\rm _V>$ ($\sigma)$}& \colhead{$<$z$>$
($\sigma)$} & \colhead{EW([OIII])} & \colhead{EW(H$\beta$)}\\
\colhead{(1)}     & \colhead{(2)} & \colhead{(3)} & \colhead{(4)} &
\colhead{(5)} & \colhead{(6)} & \colhead{(7)} & \colhead{(8)}& \colhead{(9)}}
\startdata
A3  &$<$4000  &$>$1.0   &10 (2) &   39  &--23.1 (2.1) &0.20 (0.20)  &20 (5)\tablenotemark{*} & 50 (54) \\%
A2  &$<$4000  &0.5 - 1.0& 14 (1)     &35 &--24.4 (1.6) &0.26 (0.22) &18 (9)  & 67 (63)\\%
A1  &$<$4000  &$<$0.5   &40 (12)     &74 &--24.3 (2.2) &0.26 (0.24) &44 (25) & 95 (89)\\%
B1  &4000 - 8000&   $<$0.5&  54 (28) &87 &--25.1 (1.7) &0.36 (0.24) &46 (26) & 98 (95)\\%
B1$^+$ &$>$8000  &$<$0.5   &21 (17)  &52 &--25.4 (1.4) &0.44 (0.22) &60 (34) & 99 (98)\\%
\enddata
\tablenotetext{*}{We provide the median values (in brackets) for the last two columns 
as well. Due to significantly dispersed EW distributions and the small number of 
objects in some bins, the medians may differ significantly from the averages, 
especially for [OIII].}
\end{deluxetable}

\clearpage

\begin{deluxetable} {lrrrrrrrrrrrrrrrr}
\rotate \tabletypesize{\scriptsize} \tablenum{2} \tablewidth{0pt} \tablecaption{{\sc
Line measurements}} \tablehead{ \multicolumn{1}{c}{Line Id.}&
\multicolumn{2}{c}{A3} & \multicolumn{2}{c}{A2} & \multicolumn{2}{c}{A1}  &
\multicolumn{2}{c}{B1} & \multicolumn{2}{c}{B1$^+$} \\
\cline{2-3} \cline{4-5} \cline{6-7} \cline{8-9} \cline{10-11}\\
 \colhead{} & \colhead{EW ($\sigma$)} & \colhead{FWHM ($\sigma$)} & \colhead{EW
($\sigma$)} & \colhead{FWHM ($\sigma$)}  & \colhead{EW ($\sigma$)} &
\colhead{FWHM ($\sigma$)} &\colhead{EW ($\sigma$)} & \colhead{FWHM ($\sigma$)}
& \colhead{EW ($\sigma$)} & \colhead{FWHM ($\sigma$)}}
\startdata

OVI 1034      &    -  &      -    &    - & -         &17 (4) &3770  (500)&21 (6)&4350 (600)&18 (4) &5510 (700)\\
CIII 1176     &    -  &      -    &4 (3) &3600 (2000)&1 (1)  &5100 (2000)&4 (3) &5350 (2500)& -           & - \\
Ly$\alpha$1216&48 (15)&2960 (700) &92 (21)&2960 (500)&101(25)&3450  (500)&116(21)&3700 (400)&93 (18)&5420 (400)\\
NV 1240       &26 (9)&3870 (700)  &21 (7)&4110  (700)&17 (6) &3380  (600)&28 (9)&5080 (800)&25 (6) &7250 (900)\\
SiII 1264     &3 (2)&2250  (1000) &2 (1) &1420  (800)&    -  &    -      &   -  &    -     & -   &   -       \\
SiII 1306     &6 (2)&3530  (500)  &5 (1) &2290  (300)&3 (1)  &2290  (300)&4 (1) &3670 (500)& 3 (1) &4820 (1000)\\
CII 1335      &2 (1)&1970 (500)   &3 (1) &1680  (300)&1 (1)  &2240  (300)&1 (1) &2240 (600)& 1 (1) &3590 (1500)\\
SiIV 1398     &16 (3)&4500  (400) &16 (3)&3860  (400)&13 (3) &4070  (400)&11 (2)&3640 (400)& 12 (2)&6000 (400)\\
NIV] 1486     &   -   &   -       &  -   &     -     &2 (1)  &2610  (800)&1 (1) &3020 (800)&  -    &    -     \\
CIV 1549      &23 (4)&4160  (400) &35 (9)&3290  (400)&69 (16)&3480  (400)&83 (16)&4260 (400)&79 (10)&6190 (400)\\
HeII 1640     &8 (2)&5850 (700)   &10 (2)&4750  (500)&11 (2) &4570  (400)&11 (2)&4930 (500)& 4 (1) &4750 (600)\\
OIII] 1663    &2 (1)&2880 (800)   &5 (2) &4140  (700)&4 (1)  &3060  (500)&7 (2) &4320 (700)& 5 (2) &3600 (700)\\
NIII] 1750    &4 (3) &2050 (1000) &   -  &   -       &  -    &    -      &4 (3) &4410 (1500)& -    & -        \\
AlIII 1860    &8 (4) &4670  (1200) &5 (3) &3860 (1200)&2 (1)  &4180 (1000)&3 (2) &4700 (2000)& 3 (2)&5700 (2500)\\
SiIII] 1892   &7 (3) &1900  (800) &8 (2) &1740  (700)&6 (2)  &2530  (600)&5 (3) &2530 (1000)& 3 (2)&3010 (1500)\\
CIII] 1909    &10 (4)&2350  (800) &15 (4)&1720  (700)&17 (4) &2510  (500)&20 (7)&3140 (700)& 18 (9)&3920 (1200)\\
MgII 2798     &12 (8)&2030 (700)  &14 (4)&1500  (700)&16 (2) &1600  (600)&40 (8)&3960 (600)& 44 (8)&6540 (600)\\
\feiiuv\      &22 (10)&3000   (-) &16 (8)&3000    (-)&9 (5)  &3000   (-) &7 (5) &5000 (-)   & 7 (5)&9000 (-)\\
\enddata


\end{deluxetable}

\clearpage





\begin{deluxetable}{cccccccccc}
\tabletypesize{\scriptsize}
\tablenum{3} \tablewidth{380pt}
\tablecaption{{\sc \civ\ components}}
\tablehead{ \multicolumn{1}{c}{S.T.}& \multicolumn{4}{c}{Component 1}
&& \multicolumn{4}{c}{Component 2} \\
\cline{2-5} \cline{7-10}\\
& \colhead{Type} & \colhead{$\lambda$} & \colhead{FWHM}& \colhead{Strength
($\%$)} && \colhead{Type} & \colhead{$\lambda$} & \colhead{FWHM}&
\colhead{Strength ($\%$)}}
\startdata

A3   &   Lor  &   1547.3 & 2220  &  50  &&   Blue &   1532   &  4800     & 50  \\
A2   &   Lor  &   1547.8 & 2600  &  76  &&   Blue &   1534   &  3900     & 24  \\
A1   &   Lor  &   1547.7 & 3580  & 100  &&   -    &      &       &     \\
B1   &   Gau  &   1548.3 & 4170  &  42  &&   Gau  &   1555.8 & 15590 & 58 \\
B1$^+$  &   Gau  &   1547.8 & 4780  &  28  &&   Gau  &   1552.6 & 17480 & 72 \\
\enddata
\end{deluxetable}

\begin{figure*}[htb]
 \mbox{} \vspace{15.0cm} \includegraphics{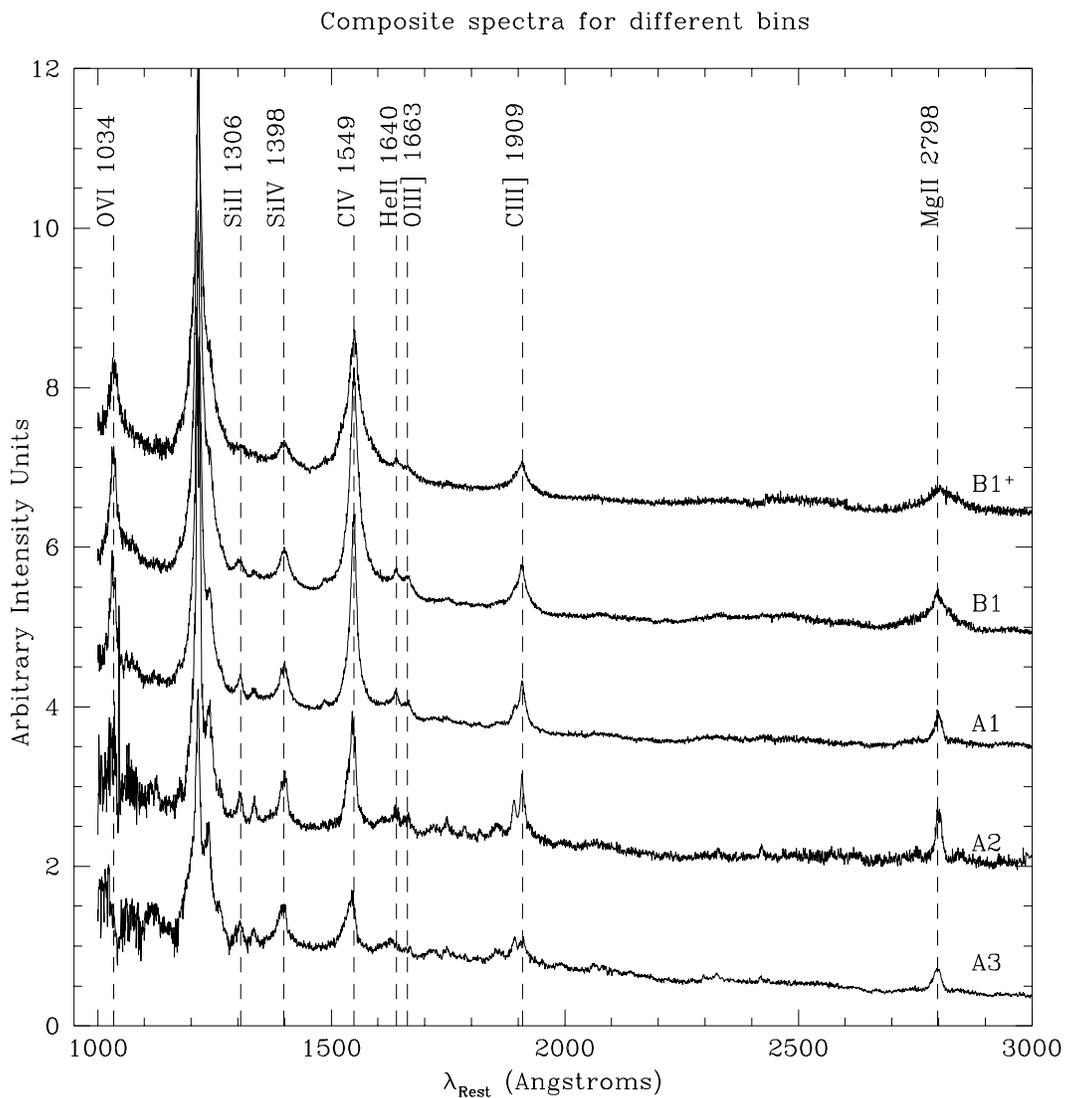} \caption[]{Median HST spectra for the 5 spectral types
 considered in this paper, with identification of the strongest emission lines.
 Each spectrum was corrected for the interstellar extinction and shifted to the rest frame before being processed.
 Horizontal scale is rest frame wavelength in \AA ngstroms,
 vertical scale is in arbitrary $F_{\rm \lambda}$ intensity units. 
 Spectra have been offset on the vertical axis for clarity.
 }
\end{figure*}

\begin{figure*}[htb]
 \mbox{}
 \vspace{15.0cm}
 \includegraphics{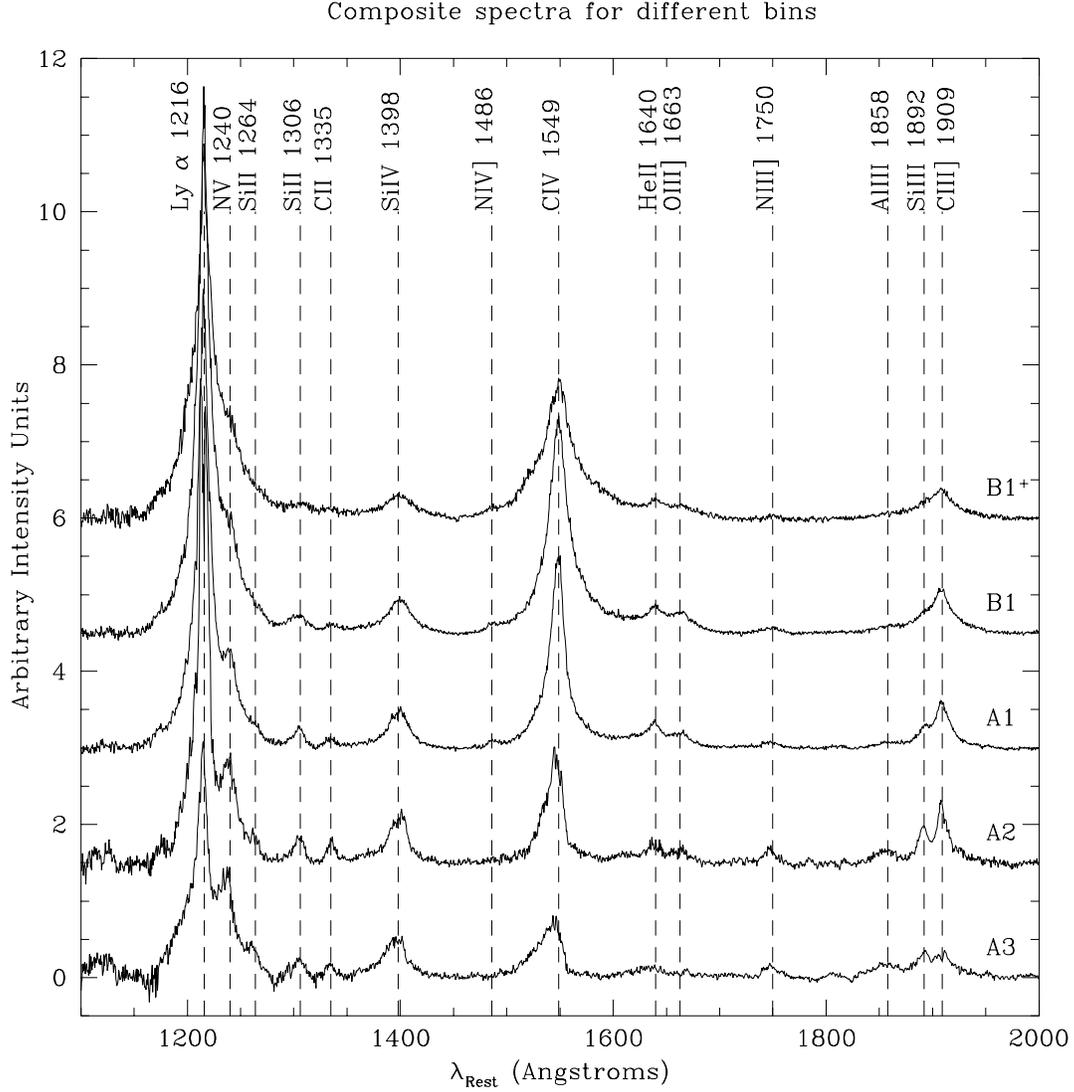}
 \caption[]{The rest frame region of the median spectra
 between 1150 \AA\ and 1950 \AA\ is shown after continuum
 and \feiiuv\ emission subtraction.
 As for Fig. 1, spectra have been offset for clarity. All lines for which we report
 equivalent width and FWHM in Table 2 are identified.
 Horizontal scale is rest frame wavelength in \AA ngstroms.}
\end{figure*}

\begin{figure*}[htb]
 \mbox{}
 \vspace{15.0cm}
 \includegraphics{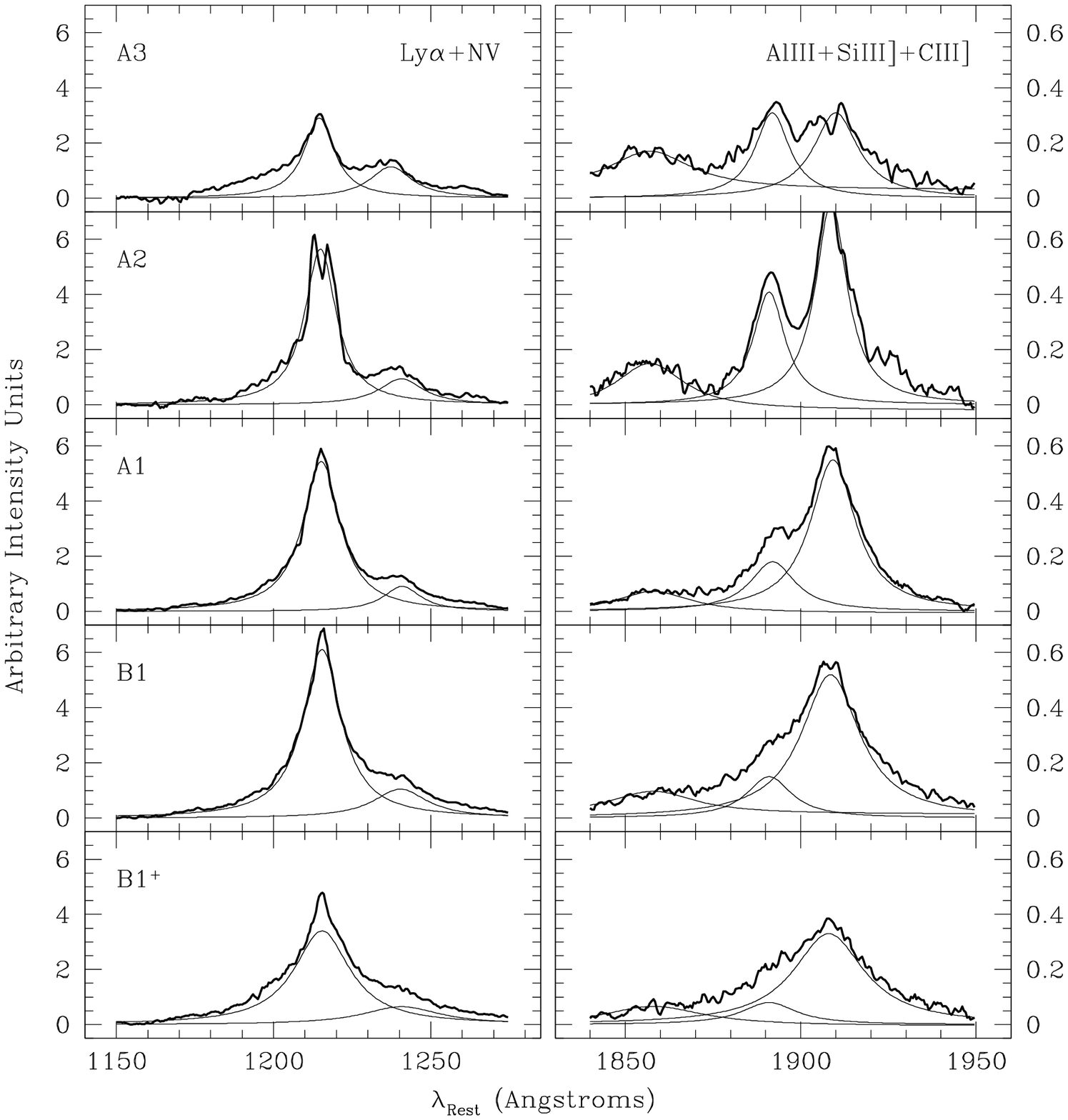}
\caption[]{Decomposition of blended lines for the spectral types
considered in this paper. Left panels: \lya\ +NV$\lambda$1240;
right panels \aliii, \siiii\ and \ciii. To all blends we subtracted in
advance the underlying continuum and an \feiiuv\ emission
template; \feiiuv\ emission is however weak around \lya\ and
\ciii. Thick lines show the profiles we fit; thin lines show
individual line contributions (assumed to have Lorentzian
profiles) to the blend. No narrow component is subtracted prior to
the fitting in any of the lines. While the NV$\lambda$1240
equivalent width remains approximately constant, the \siiii /
\ciii\ intensity ratio decreases monotonically along the sequence
of spectral types. Horizontal scale is rest frame wavelength in
\AA; vertical scale is intensity in arbitrary units.}
\end{figure*}

\begin{figure*}[htb]
 \mbox{}
 \vspace{15.0cm}
 \includegraphics{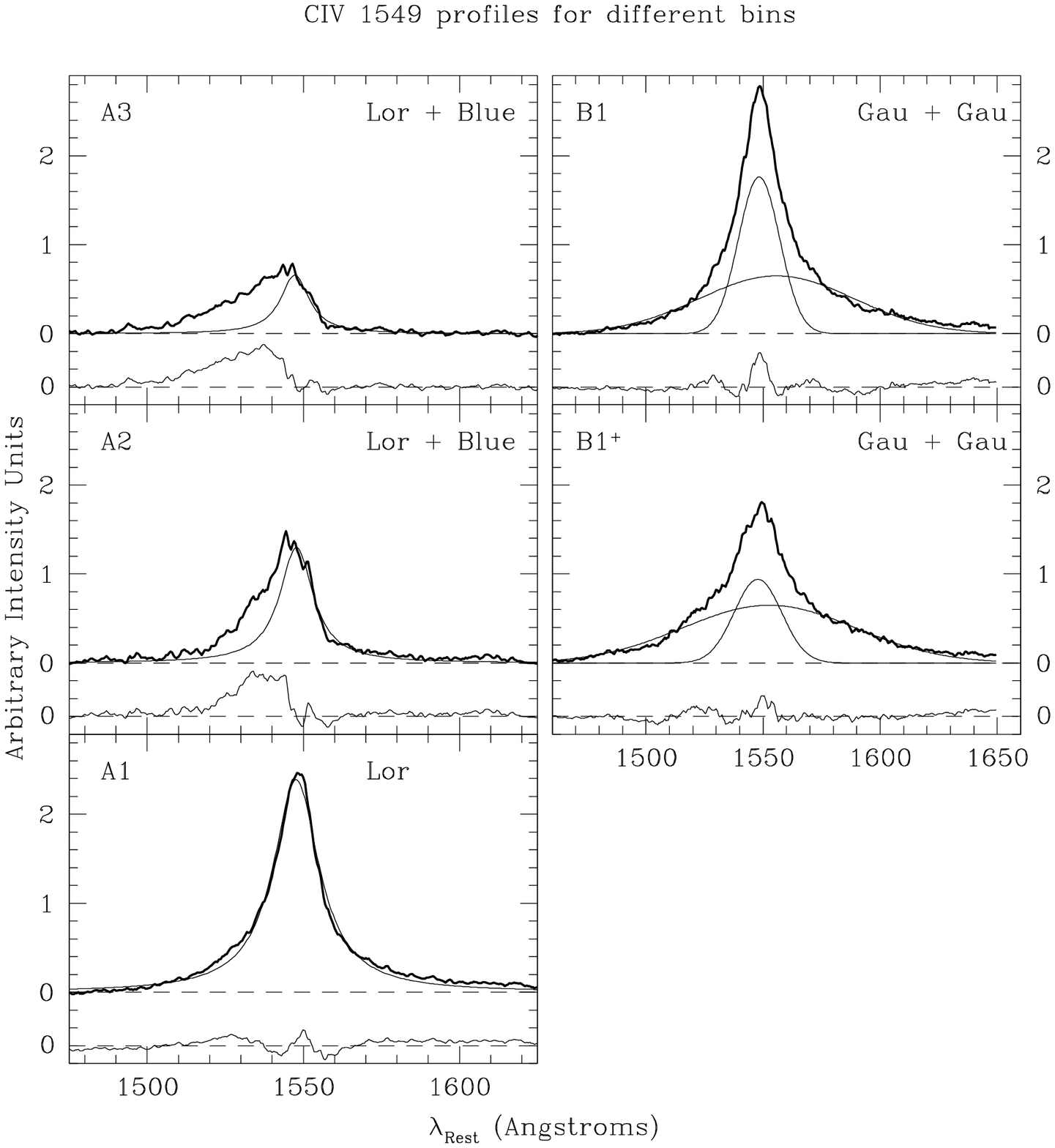}
 \caption[]{\civ\ line profile decomposition for the spectral types considered in this paper.
The fitted data are shown by thick lines; they are obtained from the continuum
and \feiiuv\ subtracted spectra after \heiiuv\ and \oiiiuv\ and NIV]$\lambda$1486
removal . For bin A3 and A2, the thin line superimposed to the data shows the
best fit provided by an unshifted Lorentzian line to the \civ\ red wing. The
residual is a broad and blueshifted feature, possibly associated with outflows.
For bin A1, a symmetric Lorentzian provides a satisfactory fit. For spectral
types B1 and B1+, two Gaussian profiles, similar to the one used by Sulentic
et al. in Paper I to fit \hbbc\ provide a satisfactory fits to the \civ\ broad
component. Narrow central residuals seen in A1 and B-bins can possibly be associated
with the contribution of the NLR.}
\end{figure*}

\begin{figure*}[htb]
 \mbox{}
 \vspace{15.0cm}
 \includegraphics{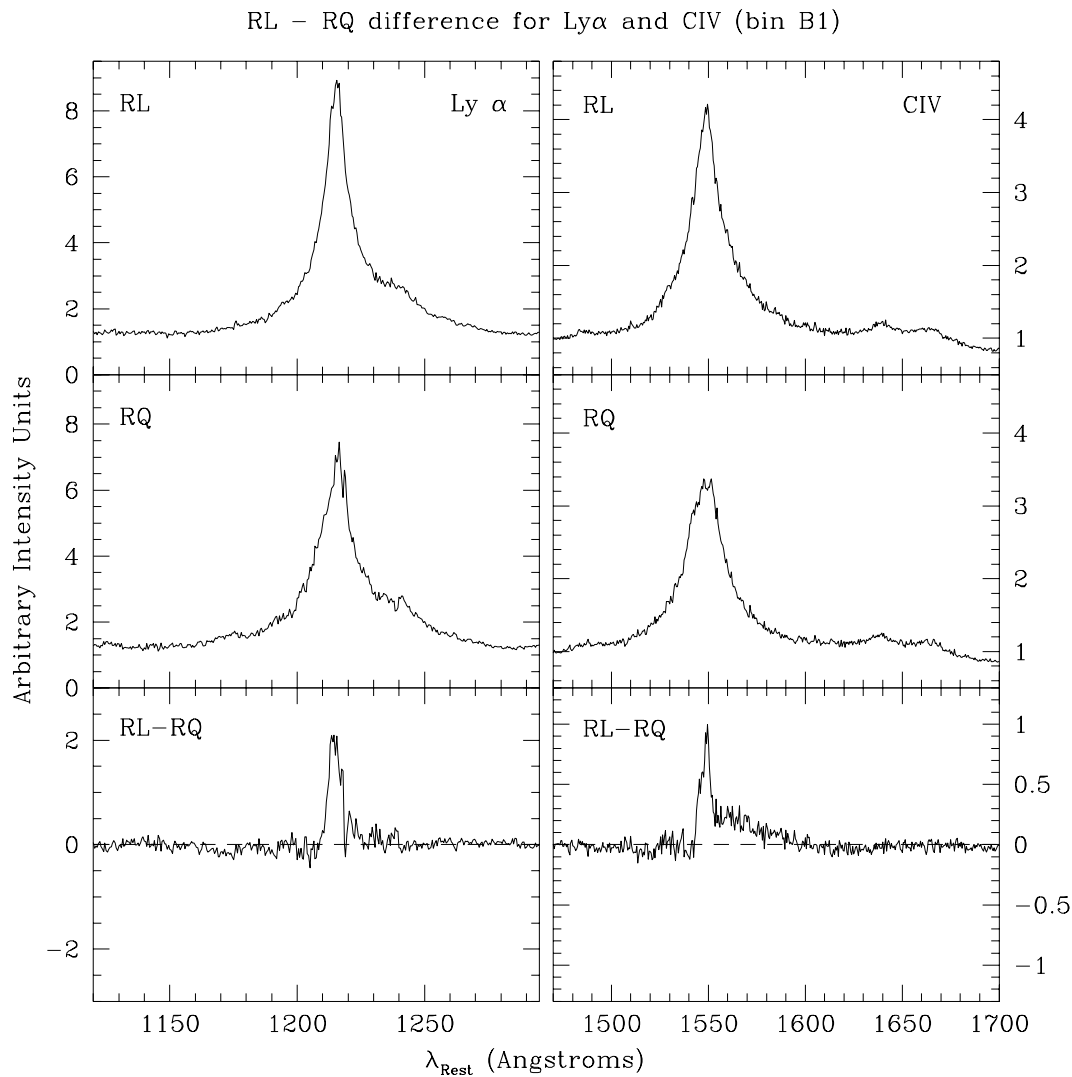}
 \caption[]{\lya\ (left panels) and \civ\ profiles (right) for radio-loud (top
panels) and radio-quiet (middle panels) quasars. Note that, unlike in Fig. 3,
 contaminating lines have not been removed. Residuals (bottom panels) show that
 the main difference is due to a stronger narrow component in radio-loud
 sources. Horizontal scale is rest frame wavelength in \AA ngstroms;
vertical scale is in arbitrary intensity units.
 }
\end{figure*}

\begin{figure*}[htb]
 \mbox{}
 \vspace{15.0cm}
 \includegraphics{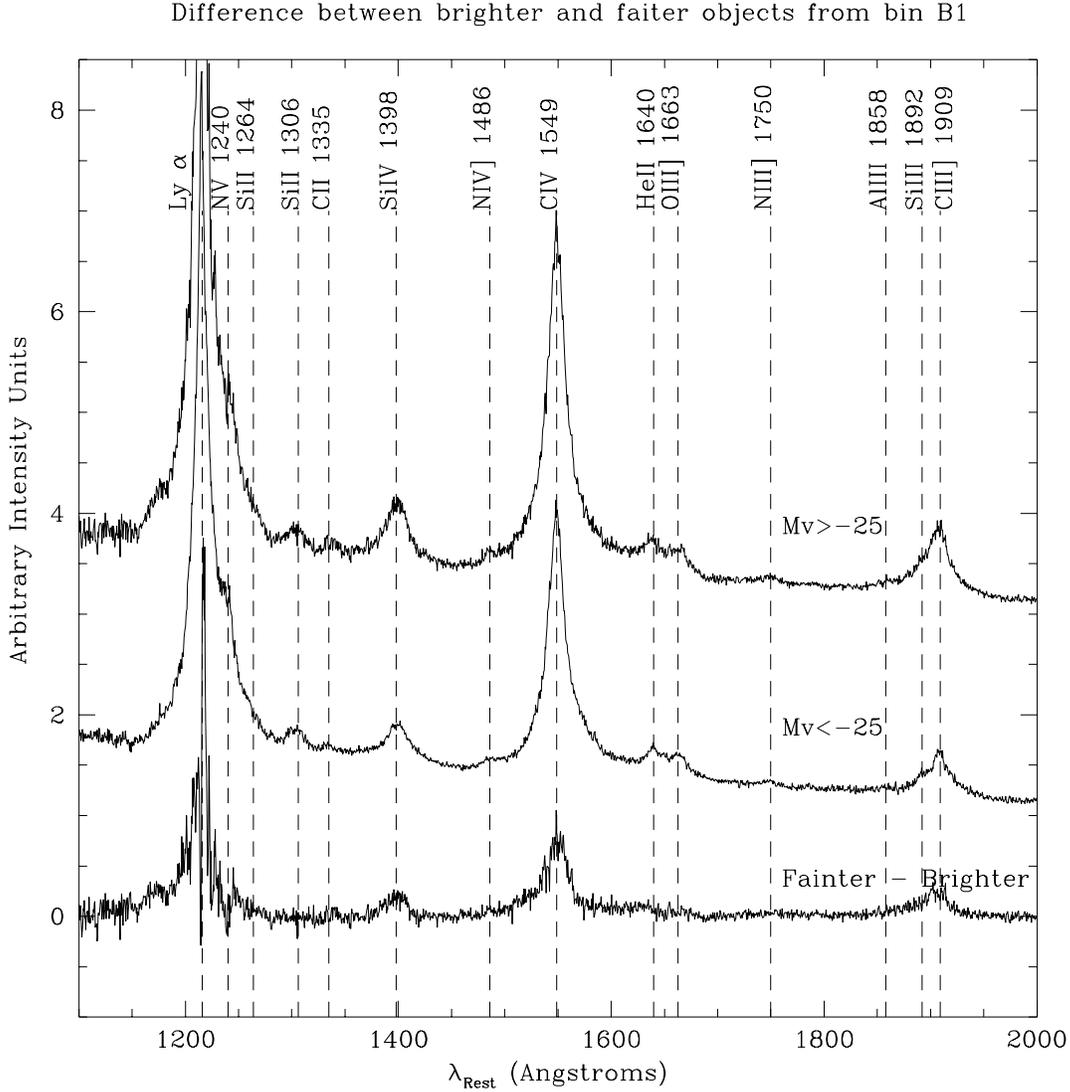}
 \caption[]{Similar division as the one shown in Fig. 5 is presented here -- this
time we show the difference between the fainter objects and the brighter objects
from bin B1. The separation value of the absolute V-band luminosity
$\rm M_V = -25$ is chosen in order to create two equally represented groups.
The Baldwin effect (anticorrelation of the equivalent width of a line with the
luminosity is apparent for lines like \lya, \siv, \civ, \siiii, but not for lines
like \nv and \siii. The situation is unclear for \heiiuv, \oiiiuv\ and other 
lines due to their lower contrast.}
\end{figure*}


\begin{thebibliography}{}
\bibitem[Axon et al.(1998)]{1998ApJ...496L..75A} Axon, D.~J., Marconi, A., Capetti, A., Maccetto, F.~D., Schreier, E., \& Robinson, A.\ 1998, \apjl, 496, L75 %
\bibitem[1999]{Bac99} Bachev R., 1999, A\&A, 348, 71
\bibitem[2004]{Bac04} Bachev R., Strigachev A., 2004, AN, 325, No. 4, 1
\bibitem[{Baskin \& Laor(2004)}]{baskin} Baskin, A.~\& Laor, A.\ 2004, \mnras, 350, L31
\bibitem[1977]{Bal77} Baldwin J. A., 1977, ApJ, 214, 679
\bibitem[1996]{Bal96} Baldwin J. A., Ferland G. J., Korista K. T, et al., 1996, ApJ, 461, 664
\bibitem[1975]{BP75} Bardeen J.M., Petterson J.A., 1975, ApJ, 195, L65
\bibitem[2003]{Ben03} Bentz M. C., Osmer P. S., 2004, AJ 127, 576
\bibitem[2002]{Bia02} Bian W., Zhao Y., 2002, A\&A, 395, 465
\bibitem[1977]{Bla77} Blandford R. D., Znajek R. L., 1977, MNRAS, 179, 433
\bibitem[1997]{Bot97} Bottorff M., Korista K. T., Shlosman I., Blandford R. D., 1997, ApJ, 479, 200
\bibitem[2001]{BG1992} Boroson T.A., Green R.F., 1992, ApJS, 80, 109
\bibitem[Boroson(2002)]{2002ApJ...565...78B} Boroson, T.~A.\ 2002, \apj, 565, 78
\bibitem[1999]{Bro99} Brotherton M. S., Francis P. J., 1999, in "Quasars and Cosmology", ASP Conference Series 162, Eds. G. Ferland and J. Baldwin.
\bibitem[Clavel et al.(1991)]{1991ApJ...366...64C} Clavel, J., et al.\ 1991, \apj, 366, 64
\bibitem[2001]{Col01} Collin S., Hur\'{e} J.-M., 2001, A\&A, 372, 50
\bibitem[2002]{Con02} Constantin A., Shields J. C., Hamann F., et al., 2002, ApJ, 565, 50
\bibitem[2003]{CS03} Constantin A., Shields U. C., 2003, PASP, 115, 592
\bibitem[1996]{Cor96} Corbin M. R., Boroson T. A., 1996, ApJS, 107, 69
\bibitem[1997]{Cor97} Corbin, M. R., 1997, ApJ, 485, 517
\bibitem[Crenshaw, Kraemer, \& Gabel(2003)]{2003AJ....126.1690C} Crenshaw, D.~M., Kraemer, S.~B., \& Gabel, J.~R.\ 2003, \aj, 126, 1690
\bibitem[2002]{Cro02} Croom S. M., Rhook K., Corbett E. A., et al., 2002, MNRAS, 337, 275
\bibitem[Czerny et al.(2003)]{2003A&A...412..317C} Czerny, B., Nikolajuk, M., 
R{\' o}{\. z}a{\' n}ska, A., Dumont, A.-M., Loska, Z., \& Zycki, P.~T.\ 2003, \aap, 412, 317
\bibitem[2004]{Cze04} Czerny B., Rozanska A., Kuraszkiewicz J., 2004, astro-ph/0403507
\bibitem[Dietrich et al.(2002)]{2002ApJ...581..912D} Dietrich, M., Hamann, F., Shields, J.~C., Constantin, A., Vestergaard, M., Chaffee, F., Foltz, C.~B., \& Junkkarinen, V.~T.\ 2002, \apj, 581, 912
\bibitem[1990]{Dum90} Dumont A. M., Collin-Souffrin S., 1990, A\&A, 229, 313
\bibitem[Francis et al.(1991)]{1991ApJ...373..465F} Francis, P.~J., Hewett, P.~C., Foltz, C.~B., Chaffee, F.~H., Weymann, R.~J., \& Morris, S.~L.\ 1991, \apj, 373, 465
\bibitem[Gammie, Shapiro, \& McKinney(2004)]{2004ApJ...602..312G} Gammie, C.~F., Shapiro, S.~L., \& McKinney, J.~C.\ 2004, \apj, 602, 312
\bibitem[2001]{Gree01} Green P. J., Forster K., Kuraszkiewicz J, 2001, ApJ, 556, 727
\bibitem[Hamann \& Ferland(1992)]{1992ApJ...391L..53H} Hamann, F.~\&
Ferland, G.\ 1992, \apjl, 391, L53
\bibitem[Hamann \& Ferland(1999)]{1999ARA&A..37..487H} Hamann, F.~\&
Ferland, G.\ 1999, \araa, 37, 487
\bibitem[Horne, Peterson, Collier, \& Netzer(2004)]{2004PASP..116..465H} Horne, K., Peterson, B.~M., Collier, S.~J., \& Netzer, H.\ 2004, \pasp, 116, 465
\bibitem[1991]{Jac91} Jackson N., Browne, I. W. A., 1991, MNRAS, 250, 422
\bibitem[Janiuk, {\. Z}ycki, \& Czerny(2000)]{2000MNRAS.314..364J} Janiuk, A., {\. Z}ycki, P.~T., \& Czerny, B.\ 2000, \mnras, 314, 364
\bibitem[2000]{Kas00} Kaspi S, Smith P. S., Netzer H., et al., 2000, ApJ, 533, 631
\bibitem[1989]{Kel89} Kellermann K. I., Sramek R., Schmidt M., Shaffer D. B., Green R., 1989, AJ, 98, 1195
\bibitem[2000]{Ker00} Kerber, F., Rosa, M., 2000, ST-ECF Newsletter No. 27, 4
\bibitem[1990]{Kin90} Kinney A. L., Rivolo A. R., Koratkar A. P., 1990, ApJ, 357, 338
\bibitem[2000]{Kom00} Komossa, S., Meerschweinchen, J., 2000, A\&A, 354, 411
\bibitem[1997]{Kor97} Korista K., Baldwin J., Ferland G., Verner D., 1997, ApJS, 108, 401
\bibitem[2004]{Kor04} Korista K. T., Goad M. R., 2004, ApJ, 606, 749
\bibitem[2001]{Kro01} Krongold Y., Dultzin-Hacyan D., Marziani P., 2001, AJ, 121, 702
\bibitem[Kuraszkiewicz et al.(2004)]{2004ApJS..150..165K} Kuraszkiewicz, J.~K., Green, P.~J., Crenshaw, D.~M., Dunn, J., Forster, K., Vestergaard, M., \& Aldcroft, T.~L.\ 2004, \apjs, 150, 165
\bibitem[2004]{Lei04} Leighly K., 2004, astro-ph/0402452
\bibitem[Maccarone, Gallo, \& Fender(2003)]{2003MNRAS.345L..19M} Maccarone, T.~J., Gallo, E., \& Fender, R.\ 2003, \mnras, 345, L19
\bibitem[1993]{Mar93} Marziani P., Sulentic J. W., Calvani M., Perez E., Moles M., Penston M. V., 1993, ApJ, 410, 56
\bibitem[1996]{M1996} Marziani, P., Sulentic, J. W., Dultzin-Hacyan, D., Calvani, M., Moles, M., 1996, ApJS, 104, 37
\bibitem[2001]{M2001} Marziani P., Sulentic J.W., Zwitter T., Dultzin-Hacyan D., Calvani M., 2001, ApJ, 558, 553 (M01)
\bibitem[2003]{Mar03a} Marziani P., Sulentic J. W., Zamanov R., Calvani M., Dultzin-Hacyan D., Bachev R., Zwitter T., 2003a, ApJS, 145, 199 (M03)
\bibitem[2003]{Mar03b} Marziani P., Zamanov R., Sulentic J. W., Calvani M., 2003b, MNRAS, 345, 1133
\bibitem[Marziani et al.(2003)]{2003MmSAI..74..492M} Marziani, P., Zamanov, R., Sulentic, J.~W., Calvani, M., \& Dultzin-Hacyan, D.\ 2003c, Memorie della Societa Astronomica Italiana, 74, 492
\bibitem[2000]{Mat00} Mathur, S., 2000, MNRAS, 314, L17
\bibitem[Mirabel(2004)]{2004astro.ph..4156M} Mirabel, I.~F.\ 2004, ArXiv Astrophysics e-prints, astro-ph/0404156
\bibitem[1997]{Mur97} Murray N., Chiang J., 1997, ApJ, 474, 91
\bibitem[Narayan, Mahadevan, \& Quataert(1998)]{1998tbha.conf..148N} Narayan, R., Mahadevan, R., \& Quataert, E.\ 1998, Theory of Black Hole Accretion Disks, edited by Marek A. Abramowicz, Gunnlaugur Bjornsson, and James E. Pringle. Cambridge University Press, 148 
\bibitem[1985]{Net85} Netzer H., 1985, MNRAS, 216, 63
\bibitem[Nicastro(2000)]{2000ApJ...530L..65N} Nicastro, F.\ 2000, \apjl, 530, L65
\bibitem[Pagani, Falomo, \& Treves(2003)]{2003ApJ...596..830P} Pagani, C., Falomo, R., \& Treves, A.\ 2003, \apj, 596, 830
\bibitem[Penston(1990)]{Pen90} Penston M. V., Croft S., Basu D., Fuller N., 1990, MNRAS,
244, 357
\bibitem[Peterson(1993)]{1993PASP..105..247P} Peterson, B.~M.\ 1993, \pasp, 105, 247
\bibitem[Peterson (1997)]{Pet97} Peterson, B. M., 1997, An introduction to active galactic nuclei, Cambridge U Press
\bibitem[Pounds, Done, \& Osborne(1995)]{1995MNRAS.277L...5P} Pounds, K.~A., Done, C., \& Osborne, J.~P.\ 1995, \mnras, 277, L5
\bibitem[Richards et al.(2002)]{2002AJ....124....1R} Richards G. T., Vanden Berk D. E., Reichard T. A., Hall P. B., Schneider D. P., SubbaRao M., Thakar A. R., York D. G., 2002, \aj, 124, 1
\bibitem[1997]{Rod97} Rodriguez-Pascual P. M., Mas-Hesse J. M., Santos-Lleo M., 1997, A\&A, 327, 72
\bibitem[2003]{Rok03} Rokaki E., Lawrence A., Economou F., Mastichiadis A., 2003, MNRAS, 340, 1298
\bibitem[2000]{Roz00} R\`{o}zanska A., Czerny B., 2000, A\&A, 360, 117
\bibitem[1998]{Sch98} Schlegel D. J., Finkbeiner D. P., Davis M., 1998, ApJ, 500, 525
\bibitem[Schneider et al.(2002)]{2002AJ....123..567S} Schneider, D.~P., et al.\ 2002, \aj, 123, 567
\bibitem[1973]{SS73} Shakura N. I., Sunyaev R. A., 1973, A\&A, 24, 337
\bibitem[Shang et al.(2003)]{2003ApJ...586...52S} Shang, Z., Wills, B.~J., Robinson, E.~L., Wills, D., Laor, A., Xie, B., \& Yuan, J.\ 2003, \apj, 586, 52
\bibitem[2004]{SG04} Snedden S. A., Gaskell, C. M., 2004, astro-ph/0403174
\bibitem[Steffen, Gomez, Raga, \& Williams(1997)]{1997ApJ...491L..73S} Steffen, W., Gomez, J.~L., Raga, A.~C., \& Williams, R.~J.~R.\ 1997, \apjl, 491, L73
\bibitem[1999]{SM99} Sulentic, J. W., Marziani, P., 1999, ApJ, 518, L9
\bibitem[Sulentic, Calvani, \& Marziani(2001)]{2001Msngr.104...25S} Sulentic, J.~W., Calvani, M., \& Marziani, P.\ 2001, The Messenger, 104, 25
\bibitem[2000]{Su2000a} Sulentic J.W., Marziani P., Dultzin-Hacyan D., 2000a, AR A\&A, 38, 521
\bibitem[2000]{Su2000b} Sulentic J. W., Zwitter T., Marziani P., Dultzin-Hacyan D., 2000b, ApJ, 536, L5.
\bibitem[Sulentic et al.(2000)]{2000ApJ...545L..15S} Sulentic J. W., Marziani P., Zwitter T., Dultzin-Hacyan D., Calvani M.\ 2000c, \apjl, 545, L15
\bibitem[Sulentic et al.(2002)]{2002ApJ...566L..71S} Sulentic J. W., Marziani P., Zamanov R., Bachev R., Calvani M., Dultzin-Hacyan D., 2002, \apjl, 566, L71 (Paper I)
\bibitem[Sulentic et al.(2004)]{2004astro.ph..5279S} Sulentic, J.~W., Stirpe, G.~M., Marziani, P., Zamanov, R., Calvani, M., \& Braito, V.\ 2004, ArXiv Astrophysics e-prints, astro-ph/0405279
\bibitem[2003]{Sul03} Sulentic J. W., Zamfir S., Marziani P., Bachev R., Calvani M., Dultzin-Hacyan D., 2003, ApJ, 597, L17
\bibitem[2003]{Tur03} Turner T. J., Kraemer S. B., Mushotzky R. F., et al., 2003, ApJ, 594, 128
\bibitem[1992]{Tyt92} Tytler D., Fan X-M, 1992, ApJS, 79, 1
\bibitem[Vanden Berk et al.(2001)]{2001AJ....122..549V} Vanden Berk, D.~E., et al.\ 2001, \aj, 122, 549 
\bibitem[2001]{vcv01} V\'eron-Cetty, M.-P., V\'eron, P.\ 2001, A\&A, 374, 92 (VCV01)
\bibitem[2001]{VVG01} V\'eron-Cetty M.-P., V\'eron P., Goncalves A. C., 2001, A\&A, 372, 730
\bibitem[1999]{Wan99} Wandel A., 1999, ApJ, 527, 649
\bibitem[2003]{War03} Warner C., Hamann F., Dietrich M., 2003, ApJ, 596, 72
\bibitem[1995]{Wil95} Wills B. J., Thompson K. L., Han M., et al., 1995, ApJ, 447, 139
\bibitem[2002]{Woo02} Woo J.-H., Urry C. M., 2002, ApJ, 579, 530
\bibitem[1999]{Xu99} Xu C., Livio M., Baum S., 1999, AJ, 118, 1169
\bibitem[2002]{ZM2002} Zamanov R., \& Marziani P., 2002, ApJ, 571, L77
\bibitem[2002]{Zam02} Zamanov R., Marziani P., Sulentic J. W., Calvani M., Dultzin-Hacyan D., Bachev R., 2002, ApJ, 567, L9






\end{thebibliography}
\end{document}